\documentclass[12pt]{article}

\usepackage{amsmath, amssymb, mathtools}
\usepackage{type1cm}
\usepackage{mathrsfs}
\usepackage{physics}
\usepackage[T1]{fontenc}                                                                                                                                                                                                                                                                                                                                                                                                                                                                                                                       
\usepackage{fancyhdr}
\usepackage{appendix}
\usepackage{titlesec}
\usepackage{cite}
\usepackage{color}
\usepackage[top=25truemm,bottom=25truemm,left=20truemm,right=20truemm]{geometry}
\usepackage[pdftex]{graphicx}

\baselineskip=\normalbaselineskip

\newcommand{\beq}{\begin{equation}}
\newcommand{\eeq}{\end{equation}}
\newcommand{\beqn}{\begin{equation*}}
\newcommand{\eeqn}{\end{equation*}}

\newcommand\als[1]{\begin{align}\begin{split}#1\end{split}\end{align}}

\makeatletter
 
 \@addtoreset{equation}{section}
\makeatother

\begin{document}
\setcounter{footnote}{0}
\setcounter{tocdepth}{3}
\bigskip
\def\thefootnote{\arabic{footnote}}

%%%%%%%%%%%%%%%%%%%%%%%%%%%%%%%%%%%%%%%%%%%%%%%%%%%
\begin{titlepage}
\renewcommand{\thefootnote}{\fnsymbol{footnote}}
\begin{normalsize}
\begin{flushright}
\begin{tabular}{l}
UTHEP-746\\
DIAS-STP-20-02
\end{tabular}
\end{flushright}
  \end{normalsize}

~~\\

\vspace*{0cm}
    \begin{Large}
%    \begin{bf}
       \begin{center}
         {The matrix regularization for Riemann surfaces with 
         magnetic fluxes}
       \end{center}
%    \end{bf}   
    \end{Large}

\vspace{0.7cm}

\begin{center}
Hiroyuki A\textsc{dachi}$^{1)}$\footnote[1]
            {
e-mail address : 
adachi@het.ph.tsukuba.ac.jp},
Goro I\textsc{shiki}$^{1),2)}$\footnote[2]
            {
e-mail address : 
ishiki@het.ph.tsukuba.ac.jp},
Takaki M\textsc{atsumoto}$^{3)}$\footnote[3]
            {
e-mail address : 
takaki@stp.dias.ie}
 and
Kaishu S\textsc{aito}$^{1)}$\footnote[4]
            {
e-mail address : 
ksaito@het.ph.tsukuba.ac.jp}

\vspace{0.7cm}

     $^{ 1)}$ {\it Graduate School of Pure and Applied Sciences, University of Tsukuba, }\\
               {\it Tsukuba, Ibaraki 305-8571, Japan}\\

     $^{ 2)}$ {\it Tomonaga Center for the History of the Universe, University of Tsukuba, }\\
               {\it Tsukuba, Ibaraki 305-8571, Japan}\\
                                  
     $^{ 3)}$ {\it School of Theoretical Physics, Dublin Institute for Advanced Studies }\\
               {\it 10 Burlington Road, Dublin 4, Ireland}\\
               \end{center}

\vspace{0.5cm}

\begin{abstract}
\noindent
We consider the matrix regularization of fields on a Riemann surface
which couple to gauge fields with a nonvanishing magnetic flux.
We show that such fields are described as rectangular matrices 
in the matrix regularization. 
We construct the matrix regularization explicitly for the case of 
the sphere and torus based on the Berezin-Toeplitz quantization, 
and also discuss a possible generalization to cases with higher genera.
We also discuss the matrix version of the Laplacian 
acting on the rectangular matrices.
\end{abstract}

\end{titlepage}

\tableofcontents

%%%%%%%%%%%%%%%%%%%%%%%%%%%%%%%%%%%%%%%%%%%%%%%%%%%
\section{Introduction}
The matrix regularization plays important roles in the matrix-model 
formulations of M-theory or superstring theory
\cite{Banks:1996vh, Ishibashi:1996xs}.
The first quantized theory of a membrane or a string 
is mapped by the matrix regularization 
to the matrix model \cite{deWit:1988wri}, 
which is conjectured to give a nonperturbative
formulation of M-theory or superstring theory.

In the matrix regularization, functions on a closed symplectic 
manifold $({\cal M}, \omega )$ are linearly mapped to $N\times N$ matrices. 
In this paper, we consider the case that the manifold ${\cal M}$ is 
a closed Riemann surface, which is relevant to the regularizations of 
closed membranes or strings.
In this case, the main property of the matrix regularization is 
that, for any $f, g \in C^{\infty}(\cal M)$,
their images $T_N(f), T_N(g) \in M_N(\mathbf{C})$ of the matrix regularization
satisfy \cite{Arnlind:2010ac}
\begin{align}
&\lim_{N\rightarrow \infty} ||T_N(f)T_N(g)-T_N(fg) || = 0,  
\nonumber\\
&\lim_{N\rightarrow \infty} ||N[T_N(f), T_N(g)]-iT_N(\{f,g \}) || = 0,
\nonumber\\
&\lim_{N\rightarrow \infty} \frac{1}{N}{\rm Tr}T_N(f) -  \frac{1}{2\pi V}\int_{\cal M} \omega f =0,
\label{properties of MR}
\end{align}
where $|| \cdot ||$ is a matrix norm,
$\{\; ,  \; \}$ is the Poisson bracket on ${\cal M}$ defined by $\omega$ and
$V= \int_{\cal M}\omega/2\pi$ is the symplectic volume.
The first two properties show that 
the matrix regularization approximately preserves 
two algebraic structures of functions associated with 
the ordinary pointwise product and the Poisson bracket.
%,
%while the third condition prohibits the trivial map with 
%$T_N(f)=0$ for all $f$.
For Riemann surfaces, the matrix regularization 
satisfying (\ref{properties of MR}) can be constructed 
by using the Berezin-Toeplitz quantization 
\cite{Klimek:1992a, Klimek:1992b, Bordemann:1993zv, Ma-Marinescu}, 
which we will review later.

In this paper, we consider a generalization of the above setup to 
Riemann surfaces with non-zero magnetic flux.
Suppose that there exists a $U(1)$ magnetic flux on ${\cal M}$ as 
$\int_{\cal M}F/2\pi = Q$ with $Q$ a non-zero integer.
Note that the gauge field $A$ of the 
field strength $F$ cannot be globally defined, since 
any globally defined connection would lead to
$\int_{\cal M} dA/2\pi = \int_{\partial M} A/2\pi = 0$ for 
a closed manifold. 
The gauge field $A$ should be defined on each 
local patch and, on an overlap of any two patches, 
they are related to each other by gauge transformations. 
A typical example is given by the Wu-Yang monopole configuration
 on $S^2$, which we will review in later sections. 
Complex scalar fields coupling to $A$ through 
the gauge covariant derivative are also defined locally and 
receive gauge transformations on the overlaps.
In the matrix regularization, only globally defined functions are 
usually considered. We will consider the matrix regularization of 
locally defined fields, which couple to the gauge field 
of the nontrivial magnetic flux. 
This setup will be relevant for describing D-branes in terms of matrices, 
on which there can exist nontrivial gauge fluxes.

The locally defined scalar fields are mathematically said to 
be local sections of the complex line bundle on ${\cal M}$ with the 
connection $A$, where the globally defined fields correspond 
to the special case of the trivial bundle with $A=0$.
The local sections form a module of the algebra $C^{\infty}(\cal M)$.
Here, a left module $M_L$ of a unital algebra ${\cal A}$ is 
an abelian group such that there exists an operation 
$\cdot :\; {\cal A}\times M_L  \rightarrow M_L$ which satisfies
\begin{align}
f \cdot (a + b) &= f\cdot a + f \cdot b, 
\nonumber\\
(f+g)\cdot a &= f \cdot a + g \cdot a,
\nonumber\\
(f g) \cdot a &= f \cdot (g \cdot a),
\nonumber\\
1_{\cal A} \cdot a &= a,
\label{condition for module}
\end{align} 
for all $f, g \in {\cal A}$ and $a, b \in M_L$, where 
$1_{\cal A}$ is the identity element of ${\cal A}$.
Similarly, the right module can also be defined with the right multiplication.
For the case of the local sections of the line bundle, 
${\cal A}=C^{\infty}({\cal M})$ and
multiplying an element 
of $C^\infty ({\cal M})$ to local sections 
gives the operation $\cdot$. 
In physical terminology, (\ref{condition for module}) is 
just the property that $U(1)$ charged fields with the same charge
form a vector space and a product of a $U(1)$ charged field 
and a neutral field gives another charged field with the same charge.
In this case, the left and right multiplication gives the same 
operation, so the local sections give 
a left and right modules of the algebra $C^\infty ({\cal M})$.

The Serre-Swan theorem \cite{Serre-Swan} 
states that vector bundles on ${\cal M}$
are dual to modules of the corresponding algebra of functions on ${\cal M}$.  
%The above case of the local sections 
%is one of examples of this duality. 
The fuzzy counterpart of this theorem would suggest a correspondence 
between the fuzzy version of vector bundles and modules of 
the matrix algebra $M_N(\mathbf{C})$. 
Any module of $M_N(\mathbf{C})$ can be written as a set of rectangular 
matrices\footnote{The set of all $N\times N'$ or $N' \times N$ matrices gives 
a left or right modules of $M_N(\mathbf{C})$, respectively.}.
Thus, it is expected that the matrix regularization should be 
generalized such that the charged scalar fields are mapped 
to rectangular matrices. 

For the fuzzy sphere, there is indeed 
such a mapping from local sections to rectangular matrices 
\cite{Grosse:1995jt,Baez:1998he,Dasgupta:2002hx} (see also 
\cite{CarowWatamura:2004ct, Dolan:2006tx} for the fuzzy $\mathbf{CP}^n$). 
In \cite{Hawkins:1997gj, Hawkins:1998nj}, it is shown that the map 
can be formally constructed for K\"{a}hler manifolds.  
The main property of this map is that the relation,
\begin{align}
\lim_{N\rightarrow \infty} ||T_N(f)T_{NN'}(a)-T_{NN'}(f\cdot a) || = 0,
\label{property of rectangular regularization}
\end{align}
holds for any smooth function $f$ and local section $a$ of a complex line bundle,
where $T_{NN'}$ is the linear map from local sections to 
$N\times N'$ rectangular matrices.
The difference $N'-N$ corresponds to the monopole 
charge (the Chern number) of the line bundle and this should 
be kept fixed when one takes the large-$N$ limit.
The property (\ref{property of rectangular regularization})
guarantees that the structure of the module (\ref{condition for module})
is approximated well in terms of the rectangular matrices.
Note that when $N=N'$, the charge is vanishing and the local sections 
are just ordinary functions. In this case, $T_{NN'}$ reduces to $T_{N}$ and
(\ref{property of rectangular regularization}) just means 
the first property of (\ref{properties of MR}).

In this paper, after presenting a general construction of 
the maps $T_N$ and $T_{NN'}$,
we first show that this construction can be embedded in 
the Berezin-Toeplitz quantization in a $U(2)$ gauge theory.
Then, we explicitly demonstrate the construction for
the sphere and the torus.  
In the case of the fuzzy sphere, this construction gives 
the well-known fuzzy spherical harmonics
\cite{Grosse:1995jt,Baez:1998he,Dasgupta:2002hx,Dolan:2006tx}. 
For the fuzzy torus, this provides 
rectangular matrices written in terms of elliptic functions.
We will also construct fuzzy versions of the Laplacians, which 
act on the rectangular matrices and realize the continuum 
spectra in the commutative limit.

We also discuss the case of 
Riemann surfaces with higher genera.
In this case, we could not explicitly
construct the mappings due to some technical difficulties.
In particular, we will discuss that obtaining the orthonormal basis 
of spinors, which is necessary for defining each matrix element 
of $T_N$ and $T_{NN'}$,
is technically difficult to compute, though a non-orthonormal basis
can be generally written down.
Nevertheless, we present a general form of the Bergman kernel,
which formally defines the map $T_{NN'}$.

This paper is organized as follows. 
In Section 2, we review the Berezin-Toeplitz quantization, which 
gives systematic constructions of $T_N$ and $T_{NN'}$.
We also discuss that these constructions are unified in 
the Berezin-Toeplitz quantization in a $U(2)$ gauge theory. 
In Section 3 and 4, we explicitly construct this mapping for the 
case of the sphere and the torus, respectively.
In Section 5, we discuss the generalization to surfaces with 
higher genera.
In Section 6, we summarize our results and discuss 
possible applications.
In the appendices, we show some details.

%%%%%%%%%%%%%%%%%%%%%%%%%%%%%%%%%%%%%%%%%%%%%%%%%%%%%%%%%%%%
\section{Berezin-Toeplitz quantization}
%%%%%%%%%%%%%%%%%%%%%%%%%%%%%%%%%%%%%%%%%%%%%%%%%%%%%%%%%%%%
In this section, we review the Berezin-Toeplitz quantization and 
its generalization to rectangular matrices.
We also show that the quantizations with square and rectangular matrices 
can be reformulated in terms of a $U(2)$ gauge theory.
In the following, we denote a closed Riemann surface by ${\cal M}$.

%%%%%%%%%%%%%%%%%%%%%%%%%%%%%%%%%%%%%%%%%%%%%%%%%%%%%%%%%%%%
\subsection{Quantization for functions}
Let us first briefly outline the Berezin-Toeplitz quantization for 
$C^{\infty}({\cal M})$.
In this quantization, one first needs to construct zero modes of 
a certain Dirac operator\footnote{Instead of the Dirac zero modes,
one can use holomorphic sections
of complex line bundles \cite{Klimek:1992a, Klimek:1992b, Bordemann:1993zv}.}.
Let $N$ be the number of independent zero modes
and $\{\psi_I | I=1,2, \cdots, N \}$ be an orthonormal basis of the 
zero modes. Then, the Berezin-Toeplitz quantization is given by 
a map 
\begin{align}
T_N(f)_{IJ} = \int_{\cal M}\omega\, \psi^{\dagger}_J \cdot f \psi_I,
\label{Toeplitz operator}
\end{align}
where $\cdot$ means the contraction of spinor indices.
This map satisfies (\ref{properties of MR}), if appropriate 
geometric quantities are used in the construction of the zero modes or 
the Dirac operator, 
as we will explain below\footnote{
The Berezin-Toeplitz quantization also naturally appears in the 
Landau problem and the problem of tachyon condensation on D-branes. 
For example, see \cite{Hasebe:2010vp,Nair:2020xzn} and
\cite{Asakawa:2001vm,Terashima:2005ic}
for these contexts, respectively.}.

More detailed setup is as follows.
Let  $(g, \omega, J)$ be a K\"{a}hler structure on ${\cal M}$, 
which is a compatible 
triple of a metric, a symplectic form and a complex 
structure. 
%We assume that the symplectic form is normalized as
%\begin{align}
%\frac{1}{2\pi} \int_{\cal M} \omega = 1.
%\end{align}
The surface ${\cal M}$ has a spin structure associated with $J$.
Let $S$ be a spinor bundle on ${\cal M}$. The fiber of 
$S$ is $\mathbf{C}^2$ and sections of $S$ are spinors with two components. 
We define a Dirac operator acting on sections of $S$ by
\begin{align}
D = i\sigma^a \theta_a^{\mu} D_\mu = i \sigma^{a}\theta_a^{\mu} 
\left( \partial_{\mu} + \frac{1}{4}\Omega_{\mu bc}\sigma^b \sigma^c
-i N A_{\mu} \right),
\label{Dirac op}
\end{align}
where $\sigma^a$ $(a=1,2)$ are Pauli matrices,  
$\Omega_{\mu ab}$ and $\theta^\mu_a$ are the 
spin connection and the inverse of the zweibein  
for the metric $g$, respectively, and 
$N$ is a positive integer corresponding to the charge of the spinor fields.
We choose the gauge field $A$ to be the symplectic potential, 
namely, $A$ is given by $\omega =  V dA  $ on each local patch, 
where $V= \int_{\mathcal{M}} \omega/2\pi$. 
The field strength $F=dA$ then satisfies $\int_{\mathcal{M}} F/2\pi = 1$.
See Appendix \ref{Dirac operator on Riemann surface}
for a detailed definition of the Dirac operator.
With this set up, it follows from the index theorem that 
the number of zero modes of (\ref{Dirac op}) is equal to $N$. 
Let $\{\psi_I | I=1,2, \cdots, N \}$ be an orthonormal basis of 
the zero modes, with respect to the inner product 
\begin{align}
(\psi, \psi') = \int_{\cal M} \omega\, \psi^\dagger \cdot \psi'.
\label{inner product of spinors}
\end{align}
Then, the Toeplitz operator for $f \in C^{\infty}({\cal M})$ is 
defined by (\ref{Toeplitz operator}). 
It is shown that with this definition, (\ref{Toeplitz operator}) 
satisfies the main properties (\ref{properties of MR}) of the 
matrix regularization
\cite{Bordemann:1993zv, Ma-Marinescu}.

%%%%%%%%%%%%%%%%%%%%%%%%%%%%%%%%%%%%%%%%%%%%%%%%%%%%%%%%%%%%%%%%%%%%%
\subsection{Quantization for local sections}
The spinor fields of $S$ transform as 
$\psi \rightarrow e^{i N \alpha }\psi $ under a gauge transformation, where
$\alpha$ is a local gauge parameter.
We can consider complex scalar fields with charge $Q$, which transform
as $\varphi \rightarrow e^{i Q \alpha }\varphi$.
In this subsection, we consider a matrix regularization of such charged
scalar fields. 

One cannot use (\ref{Toeplitz operator}) for the charged fields,
since $\psi^{\dagger}_J \cdot \varphi \psi_I$ is not gauge invariant.
In order to make a gauge invariant mapping, we introduce 
two copies of the spinor bundles $S^{(N)}$ and $S^{(N')}$
which have the same connection $A$ but different charges, $N$ and $N'$,
respectively. 
Let $\{\psi_I^{(N)}| I=1,2, \cdots, N \}$ and $\{\psi_I^{(N')} | I=1,2, \cdots, N' \}$ 
be orthonormal bases of the Dirac zero modes, which transform 
as $\psi_I^{(N)} \rightarrow e^{i N \alpha }\psi_I^{(N)}$ and 
$\psi_I^{(N')} \rightarrow e^{i N' \alpha }\psi_I^{(N')}$. 
Then, for a given charged scalar $\varphi$ with charge $Q$, 
we define a rectangular matrix 
\begin{align}
T_{NN'}(\varphi)_{IJ} = \int_{\cal M}\omega\, \psi^{(N')\dagger}_J \cdot \varphi \psi_I^{(N)}.
\label{rectangular map}
\end{align}
In order for this to be gauge invariant, $N$ and $N'$ have to be 
related by $N'-N  = Q$.
Note that (\ref{rectangular map}) reduces to 
(\ref{Toeplitz operator}) for $Q=0$.
The map (\ref{rectangular map}) 
can also be formulated in terms of the Bergman kernel, which is a 
projection operator onto the Dirac zero modes.
See Appendix~\ref{Bergman kernel} for this formulation.

The map (\ref{rectangular map}) has the property 
(\ref{property of rectangular regularization})
\cite{Hawkins:1998nj}. See Appendix~\ref{Proof of 1.3} for the 
proof.
Thus, it indeed gives a natural matrix regularization of 
the charged scalar fields.

%%%%%%%%%%%%%%%%%%%%%%%%%%%%%%%%%%%%%%%%%%%%%%%%%%%%%
\subsection{Quantization in $U(2)$ gauge theory}
\label{Quantization in U(2) gauge theory}
Here, we show that the map (\ref{rectangular map}) is naturally 
obtained as Toeplitz operators
in a $U(2)$ gauge theory. 

Note that the definition of (\ref{Toeplitz operator}) can be generalized such 
that the gauge group of $A$ is non-abelian
\cite{Nair:2020xzn,Hasebe:2017myo,Ishiki:2018aja}. 
Suppose that the gauge group is $U(2)$ and $S$ is the spinor bundle 
in the fundamental representation.
We represent the 2-dimensional vector space of the fundamental representation 
by using the two-component representation
\begin{align}
\left(
\begin{array}{c}
\psi \\
\psi' \\
\end{array}
\right),
\end{align}
where each upper and lower component is a 
spinor on ${\cal M}$.
On these spinors, we can consider actions of adjoint scalar
fields of $U(2)$, which can be represented as $2\times 2$ matrices:
\begin{align}
\left(
\begin{array}{c}
\psi \\
\psi' \\
\end{array}
\right) 
\rightarrow 
\left(
\begin{array}{cc}
\varphi_{11} & \varphi_{12} \\ 
\varphi_{21} & \varphi_{22} \\
\end{array}
\right)
\left(
\begin{array}{c}
\psi \\
\psi' \\
\end{array}
\right).
\label{action of adjoint scalar}
\end{align}
We then consider a quantization of the algebra of the adjoint scalars
using the Toeplitz operators.

In order to realize the mapping (\ref{rectangular map}), 
let us consider a special case\footnote{It is worth pointing out that
this configuration of the $U(2)$ connection is also
recovered as solutions of the equation of motion of a
matrix model which formulates pure Yang-Mills theory on
fuzzy spaces \cite{Steinacker:2003sd, Grosse:2004wm}.}
where only the connection of a diagonal $U(1)$ subgroup is
nontrivial and the Dirac operator takes the form,
\begin{align}
D =  i \sigma^{a}\theta_a^{\mu} 
\left( \partial_{\mu} + \frac{1}{4}\Omega_{\mu bc}\sigma^b \sigma^c
-i 
\left(
\begin{array}{cc}
N A_{\mu} & 0 \\
0 & N' A_{\mu} \\
\end{array}
\right)
\right),
\end{align}
where $A_{\mu}$ is a $U(1)$ connection satisfying 
$\int_{\mathcal{M}} F/2\pi =1 $. 
Namely, the upper and lower components of the fundamental representation
transform as spinors with charge $N$ and $N'$, respectively, under the 
diagonal $U(1)$ gauge transformation.
The spinor bundle is thus decomposed to a direct sum of 
$S^{(N)}$ and $S^{(N')}$ introduced in the previous subsection.
From (\ref{action of adjoint scalar}), we see that
the $(1,2)$ and $(2,1)$ elements of an adjoint scalar field 
behave as fields with charge $N-N'$ and $N'-N$, respectively, while 
the diagonal elements behave as neutral fields.
A basis of zero modes of the Dirac operator is given by 
$\{ \Psi_I | I=1,2, \cdots, N+N' \}$ with
\begin{align}
&\Psi_I =
\left(
\begin{array}{c}
\psi_I^{(N)} \\
0 \\
\end{array}
\right)   \;\;\; {\rm for} \;\; I=1,2, \cdots, N, \nonumber\\
& \Psi_I =
\left(
\begin{array}{c}
0 \\
\psi_{I-N}^{(N')} \\
\end{array}
\right)  \;\;\; {\rm for} \;\; I= N+1, N+2, \cdots, N+N' ,
\end{align}
where $\psi_I^{(N)}$ and $\psi_I^{(N')}$ are bases introduced in 
the previous subsection.
For an adjoint scalar field $\varphi$, the Toeplitz operator is defined by 
\begin{align}
T_{N+N'}(\varphi)_{IJ} = \int_{\mathcal{M}} \omega\, \Psi_J^\dagger \cdot \varphi \Psi_I.
\label{Toeplitz operator in U(2)}
\end{align}
If we consider a complex adjoint scalar with only $(2,1)$ element,
the Toeplitz operator (\ref{Toeplitz operator in U(2)}) is nonvanishing 
only for $I=1,2, \cdots, N$ and $J=N+1, N+2, \cdots, N+N'$. 
Thus, we obtain the rectangular map (\ref{rectangular map}) as a 
special case of the Berezin-Toeplitz quantization in the $U(2)$ gauge theory.
Note that the quantization of the diagonal elements of scalar fields gives
the mapping between neutral fields and square matrices. 
Hence, this formulation using the $U(2)$ gauge theory 
gives a unified quantization for charged and non-charged fields.

%The Berezin-Toeplitz quantization naturally appears in 
%systems with non-BPS D-branes. 
%The above viewpoint using the $U(2)$ gauge theory
%is relevant for systems with coincident D-branes 
%which appear from tachyon condensations.

%%%%%%%%%%%%%%%%%%%%%%%%%%%%%%%%%%%%%%%%%%%%%%%%%%%%%%%%
\section{Fuzzy sphere}
%%%%%%%%%%%%%%%%%%%%%%%%%%%%%%%%%%%%%%%%%%%%%%%%%%%%%%%%
In this section, we construct the quantization on a sphere $S^2$. 
%Since the mapping for square matrices 
%(\ref{Toeplitz operator}) can be seen as a special case of that 
%for rectangular matrices (\ref{rectangular map}), we only 
%consider (\ref{rectangular map}) in the following.

\subsection{Geometry of $S^2$ with a magnetic flux}

We define two open subsets of $S^2$
by $U_1=\{(\theta,\phi) \mid 0\leq\theta<\pi\}$ and
$U_2=\{(\theta,\phi) \mid 0<\theta\leq\pi\}$, where
$(\theta,\phi)$ are the polar coordinates and $0<\phi\le2\pi$.
We also define the stereographic coordinates on
 $U_1$ and $U_2$
by  $z=\tan(\theta/2)e^{i\phi}$ and $w=1/z$, respectively.

We define a K$\ddot{\text{a}}$hler metric on $S^2$ by
\als
{
	\label{metric and volume form on S^2}
	g=\frac{2dzd\bar z}{(1+|z|^2)^2}.
}
The compatible symplectic form is then given by
(\ref{symplectic form on Riemann surface}) and
the volume is $V=\int_{S^2}\omega/2\pi=1$.
With respect to this metric,
we can choose a K$\ddot{\text{a}}$hler potential,
which is defined by (\ref{Kahler potential}), as
\beq
	\label{Kahler potential for S^2}
	\rho=\text{log}(1+|z|^2).
\eeq
From (\ref{spin connection on Riemann surface})
and (\ref{U(1) connection on Riemann surface}),
we also choose a spin connection and $U(1)$ gauge field
defined on $U_1$ as
\als
{
	&\Omega_{12}=
	-i\frac{\bar zdz-zd\bar z}{1+|z|^2},\\
	&A=
	-\frac{i}{2}\frac{\bar zdz-zd\bar z}{1+|z|^2},
}
respectively. Note that we have
$\Omega_{ab}\sigma^a\sigma^b/4=iA\sigma_3$
in this gauge.
This gauge field is known as the Wu-Yang gauge configuration.
On the overlap region $U_1\cap U_2$,
the gauge field $A$ transforms as
\beq
	\label{gauge tra on S^2}
	A(w)=A(z)-d\phi.
\eeq

Let $\varphi^{(Q)}$ be a complex scalar field with charge $Q$ coupling to $A$.
Corresponding to the gauge transformation (\ref{gauge tra on S^2}),
$\varphi^{(Q)}$ transforms as
\beq
	\label{gauge tra of section}
	\varphi^{(Q)}(w)=e^{-iQ\phi}\varphi^{(Q)}(z),
\eeq
on $U_1\cap U_2$.
In general, $\varphi^{(Q)}$ can be expanded in terms of
the monopole harmonics $Y^{(Q)}_{lm}$ \cite{Wu:1976ge, Wu:1977qk}.
See Appendix \ref{appendix for monopole harmonics} for the
definition of $Y^{(Q)}_{lm}$.

Let $\psi=(\psi^+,\psi^-)$ be a spinor field on $S^2$,
where $\pm$ stands for the chirality of each component.
In our gauge, $\psi^\pm$ transform as
\beq
	\label{gauge tra of spinor}
	\psi^\pm(w)=e^{\pm i\phi}\psi^\pm(z),
\eeq
on $U_1\cap U_2$.
This means that $\psi^\pm$ are charged scalars with charge $\mp1$.

%\footnote{Note that the $2\times 2$ structure
%in (\ref{gauge tra of spinor}) is for the spinor components not for 
%the internal gauge group discussed in 
%section~\ref{Quantization in U(2) gauge theory}. }

%%%%%%%%%%%%%%%%%%%%%%%%%%%%%%%%%%%%%%%%%%%%%%%%%%%%%%
\subsection{Dirac zero modes on $S^2$}

We first construct a Dirac operator (\ref{Dirac op}) on $S^2$
and compute its zero modes.
The Dirac operator flips the chirality, so that it has only off-diagonal elements
in the chiral representation.
From (\ref{Kahler potential for S^2}) and (\ref{general form of Dirac operator}),
the Dirac operator is given by
\cite{Abrikosov:2002jr, Terashima:2005ic, Ishiki:2019mvq}
\als
{
	&D^+=
	\sqrt{2}i\left\{(1+|z|^2)\partial_{\bar z}+\frac{N-1}{2}z\right\},\\
	&D^-=
	\sqrt{2}i\left\{(1+|z|^2)\partial_z-\frac{N+1}{2}\bar z\right\},
}
where $D^\pm$ are the matrix elements of $D$ acting 
on the spaces with chirality $\pm 1$, respectively.
In this section, we also write $N=2J+1$ by using a half integer spin $J$.
Note that spinor fields $\psi^{(N)}=(\psi^{(N)+},\psi^{(N)-})$ on which $D$ acts
transform as
\beq
	\psi^{(N)\pm}(w)=e^{-i(N\mp1)\phi}\psi^{(N)\pm}(z).
\eeq
on $U_1\cap U_2$.
As shown below, an orthonormal basis 
$\{\psi^{(N)}_r\mid r=-J,-J+1,\ldots,J\}$ of $\text{Ker}D$
with respect to the inner product (\ref{inner product of spinors})
can be expressed in terms of the monopole harmonics
as
\beq
	\label{Dirac zero mode on S^2}
	\psi^{(N)}_r
	=(-1)^{J-r}
	\mqty(Y^{(N-1)}_{J-r}\\0).
\eeq
See appendix~\ref{appendix for monopole harmonics} for the 
definition of the monopole harmonics.

In the following, we will derive (\ref{Dirac zero mode on S^2}).
For the decomposition $\psi=(\psi^+,\psi^-)$,
the Dirac equation $D\psi=0$
reduces to two differential equations $D^\pm\psi^\pm=0$.
We can easily solve these equations and find that
$\psi^\pm=(1+|z|^2)^{\mp(N\mp1)/2}h^\pm$, where
$h^+$ and $h^-$ are arbitrary holomorphic and anti-holomorphic functions
on $U_1$, respectively. We focus on the norm of $\psi^\pm$ given by
\als
{
	\int_{S^2}\omega\,|\psi^\pm|^2=
	i\int_{S^2}dzd\bar z\, (1+|z|^2)^{\mp N-1}|h^\pm|^2.
}
We find that the norm of $\psi^-$ does not converge for $N\geq1$
unless $h^-=0$, whereas the norm of $\psi^+$ converges when
the degree of $h^+$ is less than $N$.
Therefore, we find $\text{Ker}D=\text{Ker}D^+=N$,
which is consistent with the index theorem.
Since $h^+$ can be expanded in terms of the basis
$1,z,z^2,\ldots,z^{N-1}$,
we can choose $\psi^{(N)+}_r$ as
\als
{
	\psi^{(N)+}_r
	&=\sqrt{\frac{N}{2\pi}}\frac{1}{(1+|z|^2)^J}\binom{2J}{J+r}^{1/2}z^{J-r},\\
	&=\sqrt{\frac{N}{2\pi}}\mel*{Jr}{e^{-i\theta L^{(J)}_2}}{JJ}e^{i(J-r)\phi},
}
where we used $z=\tan(\theta/2)e^{i\theta}$ in the second equality.
By comparing this with the definition of the monopole harmonics
(\ref{monopole harmonics}), we finally obtain (\ref{Dirac zero mode on S^2}).

%%%%%%%%%%%%%%%%%%%%%%%%%%%%%%%%%%%%%%%%%%%%%%%%
\subsection{Berezin-Toeplitz quantization on $S^2$}

Here, we construct the Berezin-Toeplitz map on $S^2$
and show that  (\ref{rectangular map}) relates the monopole harmonics
and the so-called fuzzy spherical harmonics.

By using these Dirac zero modes derived above, 
we can construct (\ref{rectangular map})
for a charged scalar $\varphi^{(Q)}$ as
\als
{
	T_{NN'}(\varphi^{(Q)})_{rr'}
	&=\int_{S^2}\omega\,
	\psi^{(N')\dagger}_{r'} \cdot\varphi^{(Q)} \psi^{(N)}_r,\\
	&=(-1)^{J+J'-r-r'}\int_{S^2}\omega\,
	(Y^{(N'-1)}_{J'-r'})^\ast \varphi^{(Q)} Y^{(N-1)}_{J-r},
	\label{bt for s2}
}
where $N'-N=Q$.
Since any charged scalars with charge $Q$ can be expanded in
terms of the monopole harmonics $Y^{(Q)}_{lm}$, 
we focus only on the mapping of $Y^{(Q)}_{lm}$.
We introduce a normalization as
$\tilde Y^{(Q)}_{lm}=\sqrt{2\pi}Y^{(Q)}_{lm}$
for convenience.
Then, we have
\als
{
	T_{NN'}(\tilde Y^{(Q)}_{lm})_{rr'}
	&=(-1)^{J+J'-r-r'}\sqrt{\frac{(2l+1)N}{N'}}C^{J'-r'}_{lmJ-r}C^{J'J'}_{lQ/2JJ},\\
	&=(-1)^{3J-2r-r'+l}\sqrt{N}C^{lm}_{JrJ'-r'}C^{J'J'}_{JJlQ/2}.
}
Here, we used the formula (\ref{three monopole harmonics}) in 
the first line
and the symmetric properties for the Clebsch-Gordan coefficients
in the second line.
Furthermore, by taking into account the parity\footnote{For example, $2J$ and $2r$ are not necessarily even numbers since $J$ and $r$ can take half-integers. However, $2(J+r)$ must be
an even number from $r=-J,-J+1,\ldots,J$. This also holds for the pairs in which the sign of either $J$ or $r$ or both is reversed. The similar discussion can be applied to the pairs of $J'$ and $r'$ and $l$ and $Q/2$.}
for the pairs of $J$ and $r$, $J'$and $r'$ 
and $l$ and $Q/2$ as well as the relation $2J'-2J=Q$, we find that
\beq
(-1)^{3J-2r-r'+l}=(-1)^{J-r'+l}=(-1)^{-J+r'-Q+l}=(-1)^{-J+r'-l}.
\eeq
Thus, we have
\beq
	\label{BT map of monopole harmonics}
	T_{NN'}(\tilde Y^{(Q)}_{lm})
	=(-1)^{-l}C^{J'J'}_{JJlQ/2}\hat Y_{lm(JJ')},
\eeq
where $\hat Y_{lm(JJ')}$ is the fuzzy spherical harmonics
\cite{Madore:1991bw,Grosse:1995jt,Baez:1998he,
Dolan:2006tx,Dasgupta:2002hx,Ishiki:2006yr,Ishii:2008ib}.
See Appendix \ref{appendix for fuzzy spherical harmonics} for the definition
of $\hat Y_{lm(JJ')}$.
Here, the Clebsch-Gordan coefficient 
in (\ref{BT map of monopole harmonics})
is given by
\beq
	\label{overall factor of hat Y}
	C^{J'J'}_{JJlQ/2}
	=\sqrt{\frac{(N-1)!(N+Q)!}{(N+Q/2+l)!(N+Q/2-l-1)!}}.
\eeq
From (\ref{overall factor of hat Y}),
we find that $C^{J'J'}_{JJlQ/2}=1+O(1/N)$, if $l=O(1)$ as $N\to\infty$.
This means that $T_{NN'}(\tilde Y^{(Q)}_{lm})$ coincides with
the fuzzy spherical harmonics in the large-$N$ limit  except for the 
trivial overall factor $(-1)^{-l}$.

%It follows from (\ref{BT map of monopole harmonics}) that
%$T_{NN'}(\tilde Y^{(q)}_{\alpha lm})$ satisfies (\ref{definition1 of hat Y}).
%In addition, from $\{(-1)^{-l}\}^\ast(-1)^{-l}=1$, we have
%\beq
%	\frac{1}{N}\tr \{(T_{NN'}(\tilde Y^{(q)}_{\alpha lm}))^\dagger
%	T_{NN'}(\tilde Y^{(q)}_{\alpha l'm'})\}
%	=\delta_{ll'}\delta_{mm'}(C^{J'J'}_{JJlq})^2.
%\eeq
%Note that $\{(-1)^{-l}\}^\ast(-1)^{-l}=1$ holds for both the case that
%$l$ is an integer and half-integer. Therefore, $T_{NN'}(\tilde Y^{(q)}_{\alpha lm})$
%is a basis of $M_{N\times N'}(\mathbf{C})$ and has same properties as the
%monopole harmonics.

Note that (\ref{bt for s2}) contains the ordinary matrix regularization
for functions as a special case with $N'=N$. 
For example, the Toeplitz operator for the standard embedding 
function into $R^3$,
\begin{align}
&x^1(\theta, \phi)=\sin\theta \cos \phi, \nonumber\\
&x^2(\theta, \phi)=\sin\theta \sin \phi, \nonumber\\
&x^3(\theta, \phi)=\cos\theta,
\end{align}
is given by 
\begin{align}
T_{NN}(x^A) = \frac{1}{J+1} L^{(J)A},
\end{align}
where $L^{(J)A}$ are the $(2J+1)$-dimensional representation matrices 
of the $SU(2)$ generators satisfying the Lie algebra
$[L^{(J)A}, L^{(J)B}]=i\epsilon_{ABC} L^{(J)C}$.
This is the well-known configuration for the fuzzy sphere
\cite{Madore:1991bw}.

%%%%%%%%%%%%%%%%%%%%%%%%%%%%%%%%%%%%%%%%%%%%%%%%%%%%%%%%
\subsection{Laplacian on fuzzy $S^2$}

Here, we construct the matrix Laplacian which acts on the
rectangular matrices (\ref{BT map of monopole harmonics}).

The Laplacian for functions on $S^2$
is given by the Casimir operator of $SU(2)$.
Since we have the representation of $SU(2)$ on the space
of charged scalar fields with charge $Q$ as explained in Appendix
\ref{appendix for monopole harmonics},
we can naturally define the Laplacian on the fields by
\beq
	\Delta \varphi^{(Q)}
	=
	(\mathcal{L}^{(Q)}_A)^2\varphi^{(Q)}.
\eeq
Here, $\mathcal{L}^{(Q)}_A$
are the representation of the $SU(2)$ generators defined by
(\ref{angular momentum on local section}).
By the definition (\ref{definition1 of Y}), the monopole harmonics
$Y^{(Q)}_{lm}$ are the eigenfunctions of $\Delta$ and the spectrum
is given by $\{l(l+1)\}$ where $l=|Q|/2,|Q|/2+1,\ldots,\infty$.
The only difference from the spectrum of the ordinary spherical harmonics
is the presence of the lower bound of the angular momentum $l=|Q|/2$.

Similarly, there is another representation of $SU(2)$ on the space
of rectangular matrices, which is given in Appendix
\ref{appendix for fuzzy spherical harmonics}.
From this structure, it is natural to define the matrix Laplacian by
\beq
	\hat\Delta M
	=
	(L^{(JJ')}_A\circ)^2M,
\eeq
for any $N\times N'$ matrix $M$. Here, $L^{(JJ')}_A\circ$
are the representation of the $SU(2)$ generators defined by
(\ref{angular momentum on rectangular matrix}).
By the definition (\ref{definition1 of hat Y}),
the fuzzy spherical harmonics $\hat Y_{lm(JJ')}$
are the eigenvectors of $\hat\Delta$ and the spectrum
is given by $\{l(l+1)\}$ where $l=|J-J'|, |J-J'|+1,\ldots,J+J'$.
Since we have the relation $2J'-2J=Q$,
the spectrum of $\hat\Delta$ coincides with that of $\Delta$
except for the cutoff $J+J'$, which depends on the matrix size 
and goes to infinity in the 
large-$N$ limit.
From (\ref{BT map of monopole harmonics}), we therefore have
\beq
	\hat\Delta T_{NN'}(\tilde Y^{(Q)}_{lm})
	=
	T_{NN'}(\Delta\tilde Y^{(Q)}_{lm}),
\eeq
for $l\leq J+J'$.

%%%%%%%%%%%%%%%%%%%%%%%%%%%%%%%%%%%%%%%%%%%%%%%%%%%%%%%%
\section{Fuzzy torus}
%%%%%%%%%%%%%%%%%%%%%%%%%%%%%%%%%%%%%%%%%%%%%%%%%%%%%%%%

In this section, we construct the quantization on 
a torus $T^2$.

%\cite{Cremades:2004wa, Hamada:2012wj}

%4.1-----------------------------------------------------------------------------------------------------------------------------------------------------

\subsection{Geometry of $T^2$ with a magnetic flux}
A complex torus is defined as a quotient space of the complex plane,
\begin{equation}
T^2 := \mathbf{C}/\sim, 
\label{eq:4-2-1}
\end{equation}
where $\sim$ stands for the periodic identifications of a discrete lattice:
\begin{equation}
z \sim z' \qquad \Leftrightarrow \qquad \exists n,m \in \mathbf{Z} : z = z' + n + m \tau \quad (z, z' \in \mathbf{C}).
\label{eq:4-2-4}
\end{equation}
%\begin{equation}
%w \sim w' \qquad :\Leftrightarrow \qquad \exists n,m \in \mathbf{Z} : w = w' + n \omega_1 + m \omega_2 \quad (w,w' \in \mathbf{C}),
%\label{eq:4-2-2}
%\end{equation}
%where 
%$\omega_1 $ and $ \omega_2$ 
%are two complex numbers
%satisfying $ \omega_1 / \omega_2 \notin \mathbf{R}$.
%It is useful to introduce the moduli parameter of $T^2$ as
%\begin{equation}
%\tau = \frac{\omega_2}{\omega_1} \in \mathbf{C} \setminus \{ 0 \}.
%\label{eq:4-2-3}
%\end{equation}
%which parametrizes the moduli space of the torus.
Without loss of generality, the parameter space of $\tau$ 
can be restricted to the fundamental domain with $|\tau|>1$, 
$\Im\tau > 0$ and $-1/2 <\Re\tau < 1/2$ as usual. 
%We also introduce another complex coordinate
%$z := w / \omega_1$. In this coordinate, the identification
%\eqref{eq:4-2-2} becomes
%\begin{equation}
%z \sim z' \qquad \Leftrightarrow \qquad \exists n,m \in \mathbf{Z} : z = z' + n + m \tau \quad (z, z' \in \mathbf{C}).
%\label{eq:4-2-4}
%\end{equation}
We express $z$ 
%and its complex conjugate 
in terms of two real variables $x, y$ as
\begin{equation}
%	\begin{dcases}
	z = x + \tau y,
%	\\
%	\bar{z} = x + \bar{\tau} y.
%	\end{dcases} 
\label{eq:4-2-6}
\end{equation}
where $x$ and $y$ are periodic coordinates, for 
which we take the fundamental region as $x,y \in [0,1)$.
%Then, the torus $T^2$ can also be defined as
%\begin{equation}
%T^2 = \left\{ x + \tau y \ \middle| \ x, y \in \mathbf{R} , (x, y) \sim (x, y+1) \sim (x+1, y) \right\}. \label{eq:4-2-5}
%\end{equation}

We introduce the K$\ddot{\text{a}}$hler metric on $T^2$ as
\begin{equation}
g = 2 dz d \bar{z}  \label{torus metric}.
\end{equation}
According to (\ref{symplectic form on Riemann surface}), compatible symplectic form is then given by 
\begin{equation}
\omega = i dz \wedge d\bar{z} = 2 \Im\tau \,dx \wedge dy . 
\label{eq:4-2-9}
\end{equation}
Thus, the symplectic volume will be $V= \int_{T^2} \omega/2 \pi = \Im\tau/ \pi$.
From (\ref{Kahler potential}), the K$\ddot{\text{a}}$hler potential for the metric (\ref{torus metric}) can be chosen by
\begin{equation}
\rho = | z + \zeta | ^2      \label{torus Kahler potential}
\end{equation}
where $\zeta := \zeta^1 + \tau \zeta^2 $ and
$ \zeta^1 $ and $ \zeta^2 $ are real constants 
corresponding to the gauge holonomies along the 1-cycles on 
$T^2$.
From (A.9), the $U(1)$ gauge field is given by 
\begin{equation}
	A = \frac{\pi}{\Im \tau} \Im \left[ \left( \bar{z} + \bar{\zeta} \right) dz \right].
\label{eq:4-2-12}
\end{equation}
The gauge field $A$ is periodic up to a gauge transformation as
\als
{
	\label{gauge tra of A on T2}
	&A(z+1,\bar z+1)
	=A(z,\bar z)+d\lambda_1,\\
	&A(z+\tau,\bar z+\bar\tau)
	=A(z,\bar z)+d\lambda_2,
}
where $\lambda_1$ and $\lambda_2$ are given by
the Wilson loop phases for
$x$ and $y$ directions, respectively:
\begin{equation}
	\begin{aligned}
	\lambda_1 :=& \oint dx ( A_z \partial_x z + A_{\bar{z}} \partial_x \bar{z} )
	= \frac{ \pi}{\Im \tau} \Im (z + \zeta), \\
	\lambda_2 :=& \oint dy ( A_z \partial_y z + A_{\bar{z}} \partial_y \bar{z} )
	= \frac{ \pi}{\Im \tau} \Im \left[ \bar{\tau} (z + \zeta) \right].
	\end{aligned}
\label{eq:4-2-17}
\end{equation}
The gauge transformation (\ref{gauge tra of A on T2}) is sometimes called
the twisted boundary condition.
The spin connection on $T^2$ is evidently zero from (\ref{spin connection on Riemann surface}).

We also introduce spinor fields on $T^2$.
Let $\psi^{(N)}$ be a spinor field with charge $N \in \mathbf{Z}$.
We impose the twisted boundary condition as
\begin{equation}
	\begin{aligned}
	\psi^{(N)} \left( z + 1, \bar{z} + 1 \right)
	&= \mathrm{e}^{i N \lambda_1} \psi^{(N)} \left( z, \bar{z} \right)
	=  \exp \left( \frac{i N \pi}{\Im\tau} \Im (z + \zeta) \right) \psi^{(N)}
	\left( z, \bar{z} \right) \\
	\psi^{(N)} \left( z + \tau, \bar{z} + \bar{\tau} \right)
	&= \mathrm{e}^{i N \lambda_2} \psi^{(N)} \left( z, \bar{z} \right)
	= \exp \left( \frac{i N \pi}{\Im\tau} \Im \left[ \bar{\tau} (z + \zeta) \right] \right)
	\psi^{(N)} \left( z, \bar{z} \right)
	\end{aligned}.
\label{eq:4-2-18}
\end{equation}
With this boundary condition, it is easy to see that 
covariant derivatives of $\psi^{(N)}$ also satisfy the same boundary condition.

%%%%%%%%%%%%%%%%%%%%%%%%%%%%%%%%%%%%%%%%%%%%%%%%%%%
%4.3-----------------------------------------------------------------------------------------------------------------------------------------------------

\subsection{Dirac zero modes on $T^2$}
Let us construct a Dirac operator on $T^2$.
%We first define orthonormal frame fields (zweibeins) by
%\begin{equation}
%e_A := {e_A} ^a \partial_a, \ e^A := {e^A}_a dz^a \label{eq:4-2-19}
%\end{equation}
%such that
%\begin{equation}
%e^A \left( e_B \right) = {e^A} _a {e_B} ^a = {\delta^A} _B. \label{eq:4-2-20}
%\end{equation}
%Here, the indices $A,B=1,2$ denote orthonormal frame indices.
%Up to local rotations, ${e^A}_{a}$ is given by 
%\begin{equation}
%\left( {e^A}_{a} \right):=  \frac{1}{2} \left(
%	\begin{array}{cc}
%	1 & 1 \\
%	-i & i
%	\end{array}
%	\right). \label{eq:4-2-21}
%\end{equation}
%The gamma matrices in the orthonormal frame satisfying
%$\{\Gamma_A, \Gamma_B\} = 2 \delta_{AB} \, \mathrm{id}$ 
%are given by the Pauli matrices, $\Gamma_A=\sigma_A$ 
%as in the case of $S^2$.
%We also define the gamma matrices
%in the complex coordinate $\gamma^a$ by
%\begin{equation}
%\gamma^a = g^{ab} \gamma_b = g^{ab} \Gamma_A {e^A}_b,
%\label{eq:4-2-23}
%\end{equation}
%which satisfy $\{\gamma_a, \gamma_b\} = 2 g_{ab} \, \mathrm{id}_{\mathbf{C}^2}$.
%They are explicitly given by
%\begin{equation}
%\gamma^z =  \left(
%	\begin{array}{cc}
%	0 & 2 \\
%	0 & 0
%	\end{array}
%	\right), \quad
%\gamma^{\bar{z}} =  \left(
%	\begin{array}{cc}
%	0 & 0 \\
%	2 & 0
%	\end{array}
%	\right). \label{eq:4-2-24}
%\end{equation}
%The Dirac operator $D$ is then given by 
From (\ref{torus Kahler potential}) and (\ref{general form of Dirac operator}), the Dirac operator is given by 
\als
{
	&D^+=
	\sqrt{2}i  \left\{ \partial_{\bar{z}} + \frac{N\pi}{2 \Im\tau} (z + \zeta)\right\},\\
	&D^-=
	\sqrt{2}i  \left\{ \partial_z  -  \frac{N\pi}{2 \Im\tau} (\bar{z} + \bar{\zeta}) \right\}.
}
%\begin{equation}
%\begin{aligned}
%D=& \gamma^a(\partial_a-i q A_a) \\
%=& 2 \left(
%	\begin{array}{cc}
%	0  & \partial - \frac{ N \pi}{2 \Im\tau} (\bar{z} + \bar{\zeta})  \\
%	\bar{\partial} + \frac{ N \pi}{2 \Im\tau} (z + \zeta) & 0          
%	\end{array}
%	\right)
%\end{aligned}
%\label{eq:4-2-25}
%\end{equation}
%The spin connection $\omega_a := \omega_a^{A B} \Sigma_{A B} \ 
%(\Sigma_{A B} := [ \Gamma^A, \Gamma^B ] / 4)$ vanishes because 
%$g_{ab}$ and ${e^A}_a$ are all constant:       
%\begin{equation}
%\omega_a^{AB} := {e_b}^A {\Gamma^b}_{c a} e^{c B} - e^{bB} \partial_a {e_b}^A = 0. \label{eq:4-2-26}
%\end{equation}
We decompose a Dirac spinor as $\psi^{(N)} = (\psi^{(N)+}, \psi^{(N)-})$
and we introduce $\chi^{\pm} (z, \bar{z})$ by
\begin{equation}
	\psi^{(N)\pm} (z, \bar{z})
	= \exp \left( \frac{i N \pi}{2 \Im\tau} \Im \left[ (z + \zeta)^2 \right] \right)
	\chi^{\pm} (z, \bar{z}).
	\label{eq:4-2-29}
\end{equation}
Then, the Dirac equation $D \psi^{(N)} = 0$ 
is equivalent to the following equations:
\begin{align}
&\left( \partial_{\bar{z}} +  \frac{i N \pi}{\Im\tau} \Im(z + \zeta) \right) \chi^+ = 0, \nonumber \\
&\left( \partial_z + \frac{i N \pi}{\Im\tau} \Im(z + \zeta) \right) \chi^- = 0.
 \label{eq:4-2-30}
\end{align}
The boundary conditions for $\chi^{\pm}$ are given from 
\eqref{eq:4-2-18} and \eqref{eq:4-2-29} as
\begin{align}
&\chi^{\pm} (z + 1, \bar{z} + 1) = \chi^{\pm} (z, \bar{z}), \nonumber \\
&\chi^{\pm} (z + \tau, \bar{z} + \bar{\tau}) = \exp \left( -i N \pi \mathrm{Re} \left[ \tau + 2 (z + \zeta) \right] \right) \chi^{\pm} (z, \bar{z}).
\label{eq:4-2-31}
\end{align}

Below, we solve (\ref{eq:4-2-30}) to determine $\chi^{\pm}$
\cite{Tenjinbayashi:2005sy}.
Let us first consider $\chi^+$\footnote{
Here, we will use both of $(x, y)$ and $(z,\bar{z})$ coordinates. 
We will write $\chi^+(x,y)$ or $\chi^+(z,\bar{z})$ to represent 
which coordinate we are using, but these shall be the same quantity:
$\chi^+(x,y) =\chi^+(z,\bar{z}).$}. The periodicity of $\chi^+$ along
the $x$-direction enables us to expand it in a Fourier series:
\begin{equation}
\chi^+ (x, y) = \sum_{n \in \mathbf{Z}} c_n (y) \mathrm{e}^{i 2\pi n x} .
\label{eq:4-2-32}
\end{equation}
By substituting \eqref{eq:4-2-32} into \eqref{eq:4-2-30}, 
we obtain the differential equations for the coefficients $c_n(y)$,
\begin{equation}
\frac{c'_n (y)}{c_n (y)} = i 2 \pi n \tau -2N\pi \left( \Im(\tau) y + \Im\zeta \right)  \label{eq:4-2-33}
\end{equation}
for $\forall n \in \mathbf{Z}$,
where the prime denotes the $y$-derivative.
The solution to these equations is given by
\begin{equation}
c_n (y) = k_n  \exp \left( \frac{-N\pi}{\Im\tau} \left( \Im(z+\zeta) \right)^2 \right) \exp \left( {i 2 \pi n \tau y} \right) \label{eq:4-2-34}
\end{equation}
for $\forall n \in \mathbf{Z} $, where $k_n$ are complex integration constants.
By substituting this into \eqref{eq:4-2-32}, we obtain
\begin{equation}
\chi^+ (z, \bar{z}) = \sum_{n \in \mathbf{Z}}  k_n  \exp \left( \frac{-N\pi}{\Im\tau} \left( \Im(z+\zeta) \right)^2 \right) \mathrm{e}^{i 2\pi n z} .\label{eq:4-2-35}
\end{equation}
Then, we use the boundary condition for the $y$-direction.
From the second equation in \eqref{eq:4-2-31}, we obtain 
the following recursion relation:
\begin{equation}
k_n = \exp \left( -i \pi (2n + N)\tau \right) \exp \left( -i 2 N \pi \zeta \right) k_{n+N}. \label{eq:4-2-36}
\end{equation}
The solution to this equation is 
\begin{equation}
k_n = \mathcal{N}_n \exp \left( \frac{i \pi n^2 \tau}{N} \right) \exp \left( i 2 \pi n \zeta \right). \quad (\mathcal{N}_n = \mathcal{N}_{n+N}) \label{eq:4-2-37}
\end{equation}
From \eqref{eq:4-2-29}, \eqref{eq:4-2-35} and \eqref{eq:4-2-37}, we 
obtain 
\begin{equation}
\psi^{(N)+} = \mathrm{e}^{ i N \pi (z + \zeta) \frac{\Im(z + \zeta)}{\Im\tau}} \sum_{n \in \mathbf{Z}} \mathcal{N}_n \mathrm{e}^{\frac{i \pi n^2 \tau}{N}} \mathrm{e}^{i 2n \pi (z + \zeta)}. 
\label{eq:4-2-38}
\end{equation}
Because of the condition $\mathcal{N}_n = \mathcal{N}_{n+N}$, 
$\psi^{(N)+}$ can be further decomposed into $|N|$ 
linearly independent solutions:
\begin{equation}
\psi^{(N)+} _r = \mathcal{N}_r \, \mathrm{e}^{ i N \pi (z + \zeta) \frac{\Im(z + \zeta)}{\Im\tau}} \, \vartheta
	\left[
	\begin{array}{c}
	\frac{r}{N} \\
	0
	\end{array}
	\right]
\left( N(z + \zeta), \, N \tau \right),
\label{eq:4-2-39}
\end{equation}
where 
$r=0,1,\cdots, N-1$ and
$\vartheta$ is the Jacobi-theta function defined by
\begin{equation}
\vartheta
	\left[
	\begin{array}{c}
	a \\
	b
	\end{array}
	\right]
\left( \nu, \tau \right) := \sum_{l \in \mathbf{Z}} \mathrm{e}^{i \pi (a+l)^2 \tau} \mathrm{e}^{i 2 \pi (a+l)(\nu +b)} .\label{eq:4-2-40}
\end{equation}
The negative chirality mode $\psi^{(N)-}$ can be computed in a similar way and 
is given by
\begin{align}
\psi^{(N)-} _r = \mathcal{N}_r' \, \mathrm{e}^{ i N \pi (\bar{z} + \bar{\zeta}) \frac{\Im(z + \zeta)}{\Im\tau}} \, \vartheta
	\left[
	\begin{array}{c}
	\frac{r}{N} \\
	0
	\end{array}
	\right]
\left( N(\bar{z} + \bar{\zeta}), \, N \bar{\tau} \right).
\label{psi minus for T2}
\end{align}
From the definition of Jacobi-theta function and the positivity of $\Im \tau$, $\psi^{(N)+}_r$ and $\psi^{(N)-}_r$ converge only when $N>0$ and $N<0$, 
respectively.
In the following, we assume that $N>0$, so that $\psi^{(N)-}=0$ and the zero modes
are finally given as 
\begin{align}
\psi^{(N)}_r = \left(
\begin{array}{c}
\psi_r^{(N)+} \\
0 \\
\end{array}
\right),
\label{zero modes on T2}
\end{align}
with $\psi_r^{(N)+}$ given by (\ref{eq:4-2-39}).

We can determine the normalization factor $\mathcal{N}_r$ 
in such a way that $\psi_r^{(N)}$ become orthonormal.
The inner product of the zero modes $\psi_r^{(N)}$ is computed in
Appendix~\ref{normalization calculation} and is given by
\begin{equation}
(\psi^{(N)+}_{r'}, \psi^{(N)+}_r) = \mathcal{N}_r ^2 \sqrt{2 \Im\tau/N} \delta_{r,r'}. \label{eq:4-2-45}
\end{equation}
Thus, if we put
\begin{equation}
\mathcal{N}_r =\left( \frac{N} {2 \Im\tau} \right)^{1/4}, 
\label{eq:4-2-46}
\end{equation}
$\{ \psi_r^{(N)} | r=0,1,\cdots,N-1 \}$ forms an orthonormal basis of the zero modes.

%4.4-----------------------------------------------------------------------------------------------------------------------------------------------------

\subsection{Berezin-Toeplitz quantization on $T^2$}
In this subsection, we construct the Toeplitz operator
(\ref{rectangular map}) for local sections of a complex line bundle 
on $T^2$.

To construct (\ref{rectangular map}), we need two copies of 
spinor bundles with charges $N$ and $N'$. 
Let 
$\{ \psi^{(N)}_r | r=0,1,\cdots,N-1\}$ and 
$\{ \psi^{(N')}_{r'} | r'=0,1,\cdots,N'-1\}$ be the orthonormal bases of the 
Dirac zero modes, each of which has the form (\ref{zero modes on T2}).
We consider another scalar field $\varphi^{(Q)}$ with charge $Q=N'-N$,
which satisfies the twisted boundary condition,
\begin{equation}
	\begin{aligned}
	\varphi^{(Q)} \left( z + 1, \bar{z} + 1 \right) &= \exp \left( \frac{i Q \pi}{\Im\tau} \Im (z + \zeta) \right) \varphi^{(Q)} \left( z, \bar{z} \right), \\
	\varphi^{(Q)} \left( z + \tau, \bar{z} + \bar{\tau} \right) &= \exp \left( \frac{i Q \pi}{\Im\tau} \Im \left[ \bar{\tau} (z + \zeta) \right] \right) \varphi^{(Q)} \left( z, \bar{z} \right).
	\end{aligned}
	\label{eq:4-2-55}
\end{equation}
The map (\ref{rectangular map}) 
for $\varphi^{(Q)}$ is then given by
\begin{equation}
T_{N N'} ( \varphi^{(Q)})_{r r'} := (\psi^{(N')}_{r'}, \varphi^{(Q)} \psi^{(N)}_r).
\label{eq:4-2-56}
\end{equation} 
Note that the integrand in the inner product on the right-hand side is 
gauge invariant and hence is a completely periodic function on the torus.

The map (\ref{eq:4-2-56}) reproduces the well-known configuration 
of the fuzzy torus given by the clock-shift matrices.
To see this, we consider the quantization of the functions 
$u(x,y):=\exp (i 2\pi x)$ and $v(x,y):=\exp (i 2\pi y)$, 
which are completely periodic and hence have the vanishing charge $Q=0$.
The Toeplitz operator (\ref{eq:4-2-56}) for these functions are given as
\begin{align}
U &:= {T_{N}(u)}= \mathrm{e}^{-\frac{\pi |\tau|^2}{2N \Im \tau} -i 2 \pi \zeta_1} \, S^{\dagger},  \nonumber\\
V &:= {T_{N}(v)}=\mathrm{e}^{-\frac{\pi}{2N \Im \tau} -i 2 \pi \zeta_2} \, C^{\dagger}, 
\label{eq:4-2-51}
\end{align}
where $C$ and $S$ are clock and shift matrices respectively:
\begin{equation}
C = \left(
      \begin{array}{ccccc}
	1 &   &      &          &   \\
	  & q &      &          &   \\
	  &   & q^2 &          &  \\
	  &   &      & \ddots & \\
	  &   &      &          & q^{N-1}   
	\end{array}
	\right),
\qquad
S= \left(
	\begin{array}{ccccc}
	  &   &           &    & 1 \\
	1 &   &          &    & \\
	  & 1 &          &    & \\
	  &   & \ddots &    & \\
	  &   &           & 1 & \\
	  \end{array}
    	\right) \label{eq:4-2-52}
\end{equation}
with $q=e^{i2\pi \frac{1}{N}}$. They satisfy the well-known algebra 
of the fuzzy torus\footnote{The two complex functions 
$u$ and $v$ define an embedding of $T^2$ into $\mathbf{C}^2$.
This embedding is not an isometric embedding for the metric
(\ref{torus metric}), which is in general very difficult to construct 
for general $\tau$.
However, in \cite{Schreivogl:2013qza}, a smart  
construction of such embedding and its quantization 
for special values of $\tau$ is proposed.}: $C^N = S^N =\mathrm{id}_{(N)}$ and $CS=qSC$.

Similarly, for any periodic function with $Q=0$, which can be expanded as
\begin{equation}
f(x,y)=\sum_{n,m \in \mathbf{Z}} f_{nm} \mathrm{e}^{i 2 \pi (nx + my)}, 
\label{eq:4-2-53}
\end{equation}
the Toeplitz operator takes the form of
\begin{equation}
T_{N}(f) = \sum_{n,m = 0}^{N-1} \tilde{f}_{nm} U^n V^m. 
\label{eq:4-2-54}
\end{equation}
Note that because of the relations 
$U^N, V^N \propto \mathrm{id}_{(N)}$ as well as the 
orthogonality under the trace, the matrices
$U^n V^m$ with $ n,m = 0,1,\cdots N-1 $ form  
a complete basis of $M_N (\mathbf{C})$.
The coefficients $\tilde{f}_{nm}$ in (\ref{eq:4-2-54}) are given as follows.
By a direct calculation, we can first obtain
\begin{align}
T_N(e^{2\pi i (nx+my)}) = C_{l_n, l_m, \tilde{n}, \tilde{m}}U^{\tilde{n}}
V^{\tilde{m}},
\end{align}
where $l_n$ and $\tilde{n}$ are the quotient and remainder of 
$n$ divided by $N$,
\begin{align}
n= l_n N + \tilde{n}, 
\end{align}
and $C_{l_n, l_m, \tilde{n}, \tilde{m}}$ are given by
\begin{align}
C_{l_n, l_m, \tilde{n}, \tilde{m}}
= e^{-2\pi i N(l_n \zeta^1 +l_m \zeta^2)}
e^{-\frac{\pi |\tau|}{2N \Im \tau}(n^2-\tilde{n})}
e^{\frac{\pi \bar{\tau}}{N\Im\tau}nm}
e^{-\frac{\pi }{2N \Im \tau}(m^2-\tilde{m})}
q^{\tilde{m}\tilde{n}}.
\end{align}
Then, we find that the coefficients $\tilde{f}_{nm}$ in (\ref{eq:4-2-54})
are given by
\begin{align}
\tilde{f}_{\tilde{n}\tilde{m}}= \sum_{l_n, l_m \in \mathbf{Z}} f_{nm}
C_{l_n, l_m, \tilde{n}, \tilde{m}}.
\end{align}

Now, let us consider the quantization for $Q\neq 0$.
As in the case of the sphere,
it is convenient to consider (\ref{eq:4-2-56}) 
for a basis of the local sections.
See Appendix~\ref{Orthonormal basis of local sections on the torus},
where we construct such a basis as eigenstates of the Laplacian.
In the following, we focus on
\begin{equation}
T_{N N'} ( \varphi^{(Q)}_{n,s} )_{r r'} = (\psi^{(N')}_{r'}, \varphi^{(Q)}_{n,s} \psi^{(N)}_r) \label{eq:4-2-67}
\end{equation}
for the eigenstates $\varphi^{(Q)}_{n,s}$ of the Laplacian, 
which are defined in 
Appendix~\ref{Orthonormal basis of local sections on the torus}.

For this computation, we put some useful relations in 
Appendix~\ref{Product of higher modes of Laplacian eigenstates}.
By putting $m=0$ in \eqref{eq:A-2-8},
we first obtain
\begin{equation}
\begin{aligned}
\varphi^{(Q)}_{n,s} (z, \bar{z}) \psi^{(N)+}_r (z,\bar{z})
&= \frac{1}{\sqrt{N'}} \sum_{k=0}^n \sqrt{\left(
	\begin{array}{c}
	n \\
	k
	\end{array}
	\right) \left( \frac{Q}{N'} \right)^k \left( \frac{N}{N'} \right)^{n-k}} \\
& \qquad \times \sum_{t=1}^{N'} \varphi^{(N')}_{k, r+s+Qt} (z,\bar{z}) \varphi^{(QNN')}_{n-k,Ns-Qr+QNt} (-\zeta, -\bar{\zeta}).
\end{aligned} 
\label{eq:4-2-68}
\end{equation}
By using this relation, we write (\ref{eq:4-2-67}) as
\begin{equation}
\begin{aligned}
T_{N N'} ( \varphi^{(Q)}_{n,s})_{rr'} 
&= \frac{1}{\sqrt{N'}} \sum_{k=0}^n \sqrt{\left(
	\begin{array}{c}
	n \\
	k
	\end{array}
	\right) \left( \frac{Q}{N'} \right)^k \left( \frac{N}{N'} \right)^{n-k}} \sum_{t=1}^{N'} \varphi^{(QNN')}_{n-k,Ns-Qr+QNt} (-\zeta, -\bar{\zeta}) \\
	& \hspace{7.5cm} \times (\varphi^{(N')}_{0, r'}, \varphi^{(N')}_{k,r+s+Qt}),
\end{aligned}
\label{eq:4-2-69}
\end{equation}
where the inner product in the last line is used in the sense of 
(\ref{orthonormality of local section basis on T2}).
By using (\ref{orthonormality of local section basis on T2})\footnote{
Note that the range of indices in (\ref{orthonormality of local section basis on T2})
can be extended such that
$(\varphi^{(N)}_{n,r}, \varphi^{(N)}_{m,s})= \delta_{n,m} \delta_{r,s\,(\mathrm{mod} \, N)}$.}, 
we obtain
\begin{equation}
T_{N N'} ( \varphi^{(Q)}_{n,s} )_{rr'} = \frac{1}{\sqrt{N'}} \left( \frac{N}{N'} \right)^{n/2} \sum_{t=1}^{N'} \delta_{r',r+s+Qt\,(\mathrm{mod} \, N')} \varphi^{(QNN')}_{n,Ns-Qr+QNt} (-\zeta, -\bar{\zeta}).
\label{eq:4-2-70}
\end{equation}
By substituting the analytic form (\ref{eq:4-2-66})
for $\varphi^{(Q)}_{n,r}$, we finally obtain
\begin{equation}
\begin{aligned}
T_{N N'} ( \varphi^{(Q)}_{n,s} )_{rr'} =& \frac{1}{\sqrt{2^n n!}} \left( \frac{QN}{2N' \Im\tau} \right)^{1/4} \left( \frac{N}{N'} \right)^{n/2} \sum_{t=1}^{N'} \delta_{r',r+s+Qt\,(\mathrm{mod} \, N')} \\
& \times \sum_{l \in \mathbf{Z}} H_n \left( \sqrt{2QNN' \pi \Im\tau} \left( l + \frac{Ns-Qr+QNt}{QNN'} \right) \right)  \mathrm{e}^{i QNN'\pi \left( l + \frac{Ns-Qr+QNt}{QNN'} \right)^2 \tau}.
\end{aligned}
\label{eq:4-2-71}
\end{equation} 
See also \cite{Cremades:2004wa, Hamada:2012wj}, 
in which essentially same computations are done in different contexts.

The rectangular matrices (\ref{eq:4-2-71}) are the fuzzy version
of the eigenstates of the Laplacian.
Though they look very complicated,
we will show in the following that they give approximate eigenstates of the 
matrix version of the Laplacian and the spectrum indeed agrees 
in the large-$N$ limit with that of the continuum Laplacian.

%----------------------------------------------------------------------------------------------------------------------------------------------------------
\subsection{Laplacian on fuzzy $T^2$}
In this subsection, we construct the matrix Laplacian, which acts on 
the rectangular matrices (\ref{eq:4-2-71}).
From now on, we put $\tau = i$ and $\zeta=0$ for simplicity.

We first note that there is a crucial difference between 
the spectrum of the continuum Laplacian for $Q=0$ and that for $Q\neq 0$.
When  $Q=0$, the Laplacian is given by
\begin{equation}
\begin{aligned}
\Delta & := - 2g^{ab} \partial_a \partial_b 
= -\frac{1}{\left( \Im \tau \right)^2} \left( |\tau|^2 \partial_x^2 -2 \Re\tau \partial_x \partial_y + \partial_y^2 \right) 
= - \partial_x^2 - \partial_y^2.
\end{aligned}
\label{eq:4-2-72}
\end{equation}
The spectrum of this operator is just $4\pi^2(n^2+m^2)$, where 
$n$ and $m$ are integers. For $Q\neq 0$, however, as shown in
Appendix~\ref{Product of higher modes of Laplacian eigenstates},
the spectrum becomes that of the 1-dimensional harmonic oscillator 
(or equivalently the Landau level), 
because of the relation $[D_z, D_{\bar{z}}]={\rm const.}$
As we will see below, the matrix Laplacian naturally reproduces 
both of these spectra in the large-$N$ limit.

We first consider the matrix Laplacian for $Q=0$, which is relatively
well-known. In this case, the continuum 
Laplacian can be written in terms of a Poisson bracket and 
we can construct
the matrix Laplacian by replacing the Poisson brackets with  
the commutators of matrices.
Let us introduce the Poisson bracket induced from 
the symplectic form \eqref{eq:4-2-9}:
\begin{equation}
\begin{aligned}
	\{ f,g \} :=& \omega(X_f, X_g) 
	= \frac{1}{2}(\partial_x f \partial_y g - \partial_y f \partial_x g),
\end{aligned}
\label{eq:4-2-73}
\end{equation} 
where $X_f$ is the Hamiltonian vector field of $f$, namely, it is defined by
$\omega(X_f, v)= df(v)$.
The partial derivatives can be expressed in terms of the Poisson bracket as
\begin{equation}
\begin{aligned}
	\{ \mathrm{e}^{\pm i 2\pi x} ,f\}
	= \pm \pi i \mathrm{e}^{\pm i 2\pi x} \partial_y f, \\
	\{ \mathrm{e}^{\pm i 2\pi y} ,f\}
	= \mp \pi i \mathrm{e}^{\pm i 2\pi y} 
\partial_x f.
\end{aligned}
\label{eq:4-2-74}
\end{equation}
Thus, we can express the Laplacian as
\begin{equation}
\begin{aligned}
\Delta (f) = - \frac{1}{\pi^2} \left( \{\bar{u}, \{u, f\} \} + \{\bar{v}, \{v, f\} \} \right)
\end{aligned}.
\label{eq:4-2-75}
\end{equation}
From the algebras, $\{ u, v \} = -2 \pi^2 uv$ and 
$[U,V] = (1-q)UV= -2\pi i UV/N + O(1/N^2)$,
we obtain the following mapping rule for the Poisson bracket:
\begin{equation}
T_N (\{f,g\}) = - N\pi i \left[ T_N (f),T_N (g)\right] + O(1/N) 
\label{eq:4-2-76}
\end{equation}
This suggests that a natural choice of the matrix Laplacian is 
\begin{equation}
\hat{\Delta} (F) = N^2 \left( [U^{\dagger} ,[U,F]] + [V^{\dagger} ,[V,F]] \right). \label{eq:4-2-77}
\end{equation}

From the algebra of $U$ and $V$, we can easily prove 
that $U^n V^m \, (n,m \in \mathbf{N})$ are eigenstates of 
$\hat{\Delta}$ as
\begin{equation}
\hat{\Delta} (U^n V^m) = N^2 \mathrm{e}^{-\frac{\pi}{N}} |q^{1/2}-q^{-1/2}| \left( [n]^2_q + [m]^2_q \right)U^n V^m, \label{eq:4-2-78}
\end{equation}
where 
\begin{equation}
[n]_q := \frac{q^{n/2}-q^{-n/2}}{q^{1/2}-q^{-1/2}} = \frac{\sin(n\pi/N)}{\sin(\pi/N)}.
\label{eq:4-2-79}
\end{equation}
Since $|q^{1/2}-q^{-1/2}| \to 4 \pi^2/N^2$ and 
$[n]_q \rightarrow n$ as $N\rightarrow \infty$, the spectrum 
of the matrix Laplacian reduces to
\begin{equation}
\hat{\Delta} (U^n V^m) = 4 \pi^2 \left( n^2 + m^2 \right)U^n V^m + O(1/N)
\label{eq:4-2-80}
\end{equation}
in the large-$N$ limit.
This agrees with the continuum spectrum.
%\begin{equation}
%\Delta (u^n v^m) = 4 \pi^2 \left( n^2 + m^2 \right)u^n v^m . 
%\label{eq:4-2-81}
%\end{equation}

We next construct the matrix Laplacian for $Q\neq 0$.
A natural generalization of the Laplacian 
(\ref{eq:4-2-77}) for rectangular matrices is
\begin{equation}
\hat{\Delta} (F) = N^2 \left[ U^{\dagger} \circ (U \circ F) + V^{\dagger} \circ (V \circ F) \right], 
\label{eq:4-2-82}
\end{equation}
where $F$ is an arbitrary $N\times N'$ rectangular matrix 
and the operation $\circ$ is defined by
\begin{equation}
A \circ F := A^{(N)} F - F A^{(N')}  \label{eq:4-2-83}
\end{equation}
with Toeplitz operators $A^{(N)}$ and $A^{(N')}$ with
dimension $N$ and $N'$, respectively.

\begin{figure}[b]
	\centering
	\includegraphics[width=9cm]{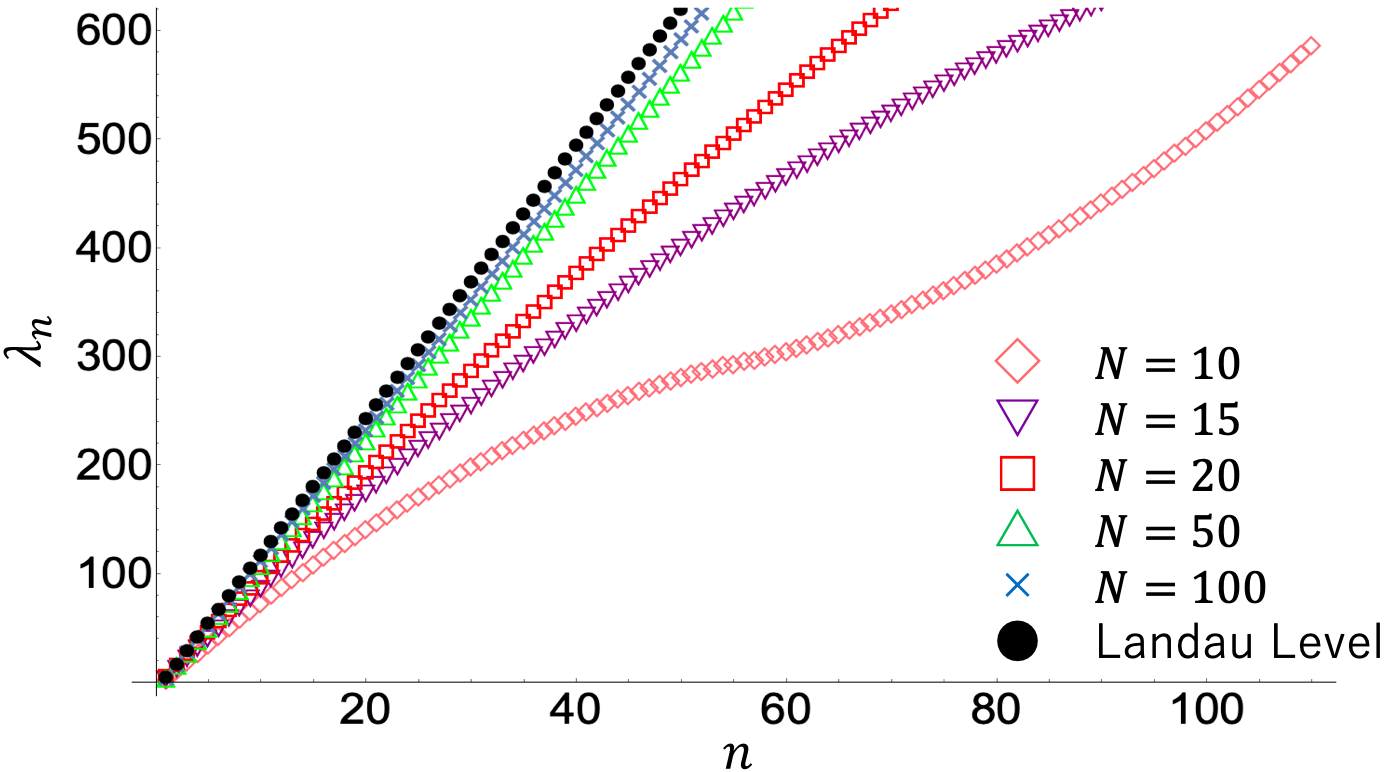}
	\caption{Numerical results of the spectrum of the Laplacian \eqref{eq:4-2-82} for finite $N \, (Q=1)$. Here $\lambda_n$ represents the $n$th smallest eigenvalue.}
	\label{fig:1}
\end{figure}

Now, let us investigate the spectrum of (\ref{eq:4-2-82}).
We compute $\hat{\Delta} (T_{NN'}(\varphi^{(Q)}_{n,s}))$, 
where $\varphi^{(Q)}_{n,s}$ is the eigenfunction of the Laplacian with charge $Q$ 
obtained in \eqref{eq:4-2-66}.
We can first show that
\begin{equation}
\begin{aligned}
\hat{\Delta} (T_{NN'}(\varphi^{(Q)}_{n,s}))_{rr'} =& 2N^2 \left( \mathrm{e}^{-\pi/N} + \mathrm{e}^{-\pi/N'} \right) T_{NN'}(\varphi^{(Q)}_{n,s}) _{rr'} \\
& - N^2 \mathrm{e}^{-\frac{\pi}{2} \left( \frac{1}{N} + \frac{1}{N'} \right)} \sum_{i=x,y \, ; \, j=\pm1} \left( \varphi^{(N')}_{0,r'} , ( \mathrm{e}^{\frac{j}{N'}D_i} \varphi^{(Q)}_{n,s}) \, (\mathrm{e}^{-\frac{jQ}{NN'}D_i} \varphi^{(N)}_{0,r} ) \right).
\end{aligned}
\label{eq:4-2-84}
\end{equation}
See Appendix~\ref{Derivation in appendix} for the derivation of
(\ref{eq:4-2-84}).
We then make an asymptotic expansion of \eqref{eq:4-2-84} in the 
large-$N$ limit as
\begin{equation}
\begin{aligned}
\hat{\Delta} (T_{NN'}(\varphi^{(Q)}_{n,s})) _{rr'} =& \left( \varphi^{(N')}_{0,r'} , \left( -D_x^2 -D_y^2\right) \varphi^{(Q)}_{n,s} \, \varphi^{(N)}_{0,r} \right) + O(1/N) \\
=& \left( \varphi^{(N')}_{0,r'} , \Delta \left( \varphi^{(Q)}_{n,s} \right) \varphi^{(N)}_{0,r} \right) + O(1/N).
\end{aligned}
\label{eq:4-2-85}
\end{equation}
This shows that the spectrum of (\ref{eq:4-2-82}) agrees 
with that in the continuum limit:
\begin{equation}
\hat{\Delta} (T_{NN'}(\varphi^{(Q)}_{n,s})) = 4Q\pi (n+1/2) T_{NN'}(\varphi^{(Q)}_{n,s}) + O(1/N).
\label{eq:4-2-86}
\end{equation}
Although we could not obtain the exact analytical solution to this eigenvalue problem for finite $N$, 
we give the numerical analysis of the spectrum of the Laplacian \eqref{eq:4-2-82}
in Fig. \ref{fig:1}.
We plotted five numerical results of the
spectrum for $N=10, 15, 20, 50, 100$ along with
the exact Landau spectrum in the case of $Q=1$.
We can see that the spectrum for finite $N$ indeed approaches to the Landau spectrum as $N$ increases.

In appendix \ref{Laplacian for rectangular matrices and Hofstadter problem},
we show that the exact eigenvalue problem of (\ref{eq:4-2-82}) 
can be mapped to the so-called Hofstadter problem \cite{1}.
Numerical studies of this problem also show that 
the eigenvalues are given as (\ref{eq:4-2-86}).

Note that if we write $U=X_1+iX_2$ and $V=X_3+iX_4$ with the 
four Hermitian matrices $X_i$ corresponding to an embedding into $R^4$, 
(\ref{eq:4-2-77}) is proportional to $[X_i, [X_i, F]]$, which 
is the natural Laplacian appearing in the matrix models.
The matrix Laplacian (\ref{eq:4-2-82}) for rectangular matrices 
also naturally appears in
the matrix models. For example, 
let us consider a block diagonal matrix configuration
in the matrix models,
\begin{align}
X_i =
\left(
\begin{array}{cc}
X_i^{(N)} & 0 \\
0 & X_i^{(N')} \\
\end{array}
\right),
\end{align}
where $X_i^{(N)}$ and $X_i^{(N')}$ are 
the configurations of the fuzzy torus with size $N$ and $N'$,
respectively.
Then, the Laplacian $[X_i, [X_i, F]]$ in the matrix models reduces to
(\ref{eq:4-2-82}) for the off-diagonal blocks of $F$, while 
it reduces to (\ref{eq:4-2-77}) for the diagonal blocks.
Thus, (\ref{eq:4-2-82}) can be seen as the Laplacian 
of the open string modes connecting the two fuzzy tori.
In fact, the same structure can also be found for the case of the sphere
\cite{Ishiki:2006yr, Ishii:2008ib}. 
It is interesting that the natural matrix Laplacian
$[X_i, [X_i, F]]$ reproduces in a unified way 
the both spectra of the charged and non-charged
fields.
%, though the two spectra are very different.

%%%%%%%%%%%%%%%%%%%%%%%%%%%%%%%%%%%%%%%%%%%%%%%%%%%%%%%%
\section{Discussion on fuzzy Riemann surfaces with higher genera}
%%%%%%%%%%%%%%%%%%%%%%%%%%%%%%%%%%%%%%%%%%%%%%%%%%%%%%%%

In this section, we discuss cases with higher genera.

%%%%%%%%%%%%%%%%%%%%%%%%%%%%%%%%%%%%%%%%%%%%%%%%
\subsection{Construction of Riemann surfaces with higher genera}
Any Riemann surface with the genus greater than $1$ can be 
constructed as the Poincar$\acute{\text{e}}$ 
disk with some identifications imposed.   
We first review this construction.
See e.g. \cite{Stillwell:1992} for more details.

Let us consider a unit disk
$D^2=\{z\in\mathbf{C}\mid|z|<1\}$ on 
the complex plane.
We adopt the Poincar$\acute{\text{e}}$ metric on $D^2$,
\beq
	\label{Poincare metric}
	g
	=
	\frac{2dzd\bar z}{(1-|z|^2)^2},
\eeq
which is the K$\ddot{\text{a}}$hler metric
compatible with the standard complex structure on 
$D^2$. The space $(D^2, g)$ is called the 
Poincar$\acute{\text{e}}$ disk.

We consider a group $SU(1,1)$, which acts on 
$D^2$ as 
\beq
	\label{linear fractional tra}
	\gamma(z)
	=
	\frac{\bar az+\bar b}{bz+a},
\eeq
where  $a$, $b$ are complex numbers satisfying 
$|a|^2-|b|^2=1$ and $\gamma$ represents an element of $SU(1,1)$,
\beq
	\label{SU(1,1) element}
	\gamma
	=
	\left(\begin{array}{cc}\bar a&\bar b\\b&a\end{array}\right).
\eeq
For any $\gamma\in SU(1,1)$, the map
$z\mapsto\gamma(z)$ is an automorphism on 
$D^2$ preserving the K$\ddot{\text{a}}$hler structure.
Note that $\gamma$ and $-\gamma$ give the same 
transformation on $D^2$, so the 
automorphism group is isomorphic to $PSU(1,1)=SU(1,1)/\mathbf{Z}_2$.
Let $\Gamma$ be a Fuchsian group, which means a 
discrete subgroup of $PSU(1,1)$.
Compact Riemann surfaces with genera greater than 1 are 
known to be constructed as a coset space $\mathcal{M}=D^2/\Gamma$.
%\footnote{Alternatively, they can also be constructed as
%coset spaces of the upper half plane 
%$\mathbf{H}=\{z\in\mathbf{C}\mid\text{Im}z>0\}$ and 
%discrete subgroups of $PSL(\mathbf{R})$, where 
%$PSL(\mathbf{R})$ is the automorphism group of 
%$\mathbf{H}$ preserving the Poincar$\acute{\text{e}}$ 
%metric $g=dzd\bar z/(\text{Im}z)^2$ on $\mathbf{H}$.}.
In this construction, all the information about the genus or the moduli 
of $\mathcal{M}$ is contained in $\Gamma$,
and $\mathcal{M}$ is given by a set of all orbits on 
$D^2$ with respect to actions of $\Gamma$. 
By analogy with the torus,
it is also useful to regard $\mathcal{M}$ as 
the disk $D^2$ with a nontrivial boundary condition 
imposed by actions of $\Gamma$.
Note that this construction also gives a natural metric on 
$\mathcal{M}$. Since $\Gamma$ preserves $g$, 
the metric (\ref{Poincare metric}) also gives 
a local K$\ddot{\text{a}}$hler metric on $\mathcal{M}$.

%%%%%%%%%%%%%%%%%%%%%%%%%%%%%%%%%%%%%%%%%%%%%%%%
\subsection{Geometric structures on $\mathcal{M}$}
We next consider charged scalars and spinor fields on $\mathcal{M}$.
For simplicity, we assume that the symplectic volume is
$V=\int_\mathcal{M}\omega/2\pi=1$,
where $\omega$ is defined by (\ref{symplectic form on Riemann surface}).

From (\ref{Poincare metric}),
a K$\ddot{\text{a}}$hler potential on $\mathcal{M}$ is 
given by
\beq
	\label{Kahler potential on higher genus}
	\rho
	=
	-\text{log}\,(1-|z^2|).
\eeq
Then, from (\ref{spin connection on Riemann surface})
and (\ref{U(1) connection on Riemann surface}),
a spin connection and $U(1)$ gauge field are
\als
{
	&\Omega_{12}=
	i\frac{\bar zdz-zd\bar z}{1-|z|^2},\\
	&A=
	-\frac{i}{2}\frac{\bar zdz-zd\bar z}{1-|z|^2}.
}
In this gauge, we have
$\Omega^{ab}\sigma_a\sigma_b/4=-iA\sigma^3$.
For $\gamma\in\Gamma$ given by (\ref{SU(1,1) element}),
the gauge field $A$ transforms as
\beq
	\label{gauge tra of gauge connection on higher genus}
	A(\gamma(z),\bar\gamma(z))
	=
	A(z,\bar z)
	+d\left(-\frac{i}{2}\text{log}\,\frac{bz+a}{\bar b\bar z+\bar a}\right).
\eeq
This is analogous to (\ref{gauge tra of A on T2}) on $T^2$.

Let $\varphi^{(Q)}$ be a complex scalar field coupling to $A$ with charge $Q$
and $\psi=(\psi^+,\psi^-)$ a spinor field on $\mathcal{M}$.
For $\gamma\in\Gamma$, they transform as
\als
{
	\label{gauge tra of section on higher genus}
	&\varphi^{(Q)}(\gamma(z),\bar\gamma(z))=
	\left(\frac{bz+a}{\bar b\bar z+\bar a}\right)^{Q/2}\varphi^{(Q)}(z,\bar z),\\
	&\psi^\pm(\gamma(z),\bar\gamma(z))=
	\left(\frac{bz+a}{\bar b\bar z+\bar a}\right)^{\pm1/2}
	\psi^\pm(z,\bar z).
}
Note that $\psi^\pm$ behave as charged scalars with charge $\pm1$.

%%%%%%%%%%%%%%%%%%%%%%%%%%%%%%%%%%%%%%%%%%%%%%%%
\subsection{Dirac zero modes and automorphic forms}

We next construct Dirac zero modes on $\mathcal{M}$.

In our case, from (\ref{Kahler potential on higher genus}) and 
(\ref{general form of Dirac operator}),
the Dirac operator is locally given by
\als
{
	&D^+=
	\sqrt{2}i\left\{(1-|z|^2)\partial_{\bar z}+\frac{N+1}{2}z\right\},\\
	&D^-=
	\sqrt{2}i\left\{(1-|z|^2)\partial_z-\frac{N-1}{2}\bar z\right\}.
}
We find that the zero modes $\psi^{(N)}=(\psi^{(N)+},\psi^{(N)-})$
take the following form:
\beq
	\label{Dirac zero mode on higher genus 1}
	\psi^{(N)\pm}
	=
	(1-|z|^2)^{(1\pm N)/2}h^\pm,
\eeq
where $h^+$ and $h^-$ are holomorphic and anti-holomorphic functions.
As we discussed in the previous subsection,
their gauge transformations are given by
\beq
	\psi^{(N)\pm}(\gamma(z),\bar\gamma(z))
	=
	\left(\frac{bz+a}{\bar b\bar z+\bar a}\right)^{(N\pm1)/2}\psi^{(N)\pm}(z,\bar z).
\eeq
By substituting (\ref{Dirac zero mode on higher genus 1}), 
we obtain the transformation of $h^\pm$ as
\als
{
	\label{tra of automorphic form}
	h^+(\gamma(z))
	&=
	(bz+a)^{(1+N)}h^+(z)
	=
	\left(\frac{d\gamma(z)}{dz}\right)^{-(1+N)/2}h^+(z),\\
	h^-(\gamma(z))
	&=
	(\bar b\bar z+\bar a)^{(1-N)}h^-(\bar z)
	=
	\left(\frac{d\bar{\gamma}(z)}{d\bar z}\right)^{-(1-N)/2}h^-(\bar z).
}
Thus, $h^\pm$ are given by automorphic forms\footnote{
Functions with the same transformation law as $h^+$
are generally called automorphic forms with weight $1+ N$.}.
The automorphic forms $h^\pm$ can be 
represented in terms of the Poincar$\acute{\text{e}}$ series
\cite{Klimek:1992b, Kra:1984} as
\als
{
	\label{automorphic form}
	h^+(z)
	&=
	\sum_{\gamma\in\Gamma}
	\left(\frac{d\gamma(z)}{dz}\right)^{(1+N)/2}f^+(\gamma(z)),\\
	h^-(\bar z)
	&=
	\sum_{\gamma\in\Gamma}
	\left(\frac{d\bar\gamma(z)}{d\bar z}\right)^{(1-N)/2}f^-(\bar\gamma(z)).
}
Here, the summations are taken over all elements of $\Gamma$ and 
$f^+$ and $f^-$ are arbitrary holomorphic and anti-holomorphic 
functions on $D^2$, respectively.
Note that for any $\gamma'\in\Gamma$, we have
\als
{
	h^+(\gamma'(z))
	&=
	\sum_{\gamma\in\Gamma}
	\left\{\frac{d(\gamma\gamma')(z)}{d\gamma'(z)}\right\}^{(1+N)/2}
	f^+((\gamma\gamma')(z))\\
	&=\sum_{\gamma\in\Gamma}
	\left(\frac{dz}{d\gamma'(z)}\right)^{(1+N)/2}
	\left(\frac{d(\gamma\gamma')(z)}{dz}\right)^{(1+N)/2}
	f^+((\gamma\gamma')(z))\\
	&=
	\left(\frac{d\gamma'(z)}{dz}\right)^{-(1+N)/2}h^+(z).
}
Thus,  (\ref{automorphic form}) indeed satisfies 
the transformation law (\ref{tra of automorphic form}).
As shown in Appendix~\ref{norm on higher genus},
the norm of $\psi^-$ does not converge.
Hence, any Dirac zero mode takes the form,
\beq
	\label{Dirac zero mode on higher genus 2}
	\psi^{(N)}(z,\bar z)
	=
	(1-|z|^2)^{(N+1)/2}
	\left(\begin{array}{cc}
	h^+(z)\\0
	\end{array}\right).
\eeq

In order to construct (\ref{Toeplitz operator}) and
(\ref{rectangular map}), we need to construct an 
orthonormal basis of the zero modes by choosing 
$h^+$ (or $f^+$) in (\ref{Dirac zero mode on higher genus 2}) appropriately.
This should be done case-by-case, since it highly depends on 
the structure of $\Gamma$.
For example, for the Bolza surface \cite{Bolza:1887, Aurich:1990vq},
which is the simplest example
of surfaces with genus 2, 
the structure of $\Gamma$ is relatively well-studied
and it might be possible to obtain an orthonormal basis 
in this case.
However, this is in general very difficult 
and is beyond the scope of this paper
(see \cite{Kra:1984} for a formal construction)\footnote{
See also \cite{Arnlind:2006ux, Arnlind:2009b, Arnlind:2019whd}
for another approach to construct the
matrix regularization of Riemann surfaces.}.

%%%%%%%%%%%%%%%%%%%%
\subsection{Quantization on ${\cal M}$}

As explained in Appendix~\ref{Bergman kernel}, the 
quantization map can be defined as (\ref{def using projection})
in terms of the projection operators
onto zero modes of the Dirac operator. 
This definition does not explicitly depends on the 
 orthonormal basis of zero modes, which is only needed
 to write down the matrix elements of (\ref{def using projection}).
Here, we show that (\ref{def using projection}) can be constructed 
for the Riemann surfaces considered above.
 
The projection is given by
the Bergman kernel (\ref{def of Bergman kernel}).
From (\ref{Dirac zero mode on higher genus 2}),
one finds that the Bergman Kernel is given by\cite{Klimek:1992b}
\als
{
	K^{(N)}(z,w)
	&=\frac{N}{2\pi}
	(1-|z|^2)^{(N+1)/2}(1-|w|^2)^{(N+1)/2}\\
	&\qquad\qquad\sum_{\gamma\in\Gamma}
	\left(\frac{d\gamma(z)}{dz}\right)^{(N+1)/2}
	(1-\gamma(z)\bar w)^{-(N+1)}
	\left(\begin{array}{cc}
	1&0\\0&0
	\end{array}\right).
	\label{BK for higher genera}
}
This is very similar to the Bergman kernel on $D^2$, 
which we review in Appendix~\ref{Bergman kernel on disk},
except that the factor $(1-z\bar w)^{-(1+N)}$ is now given 
as a Poincar$\acute{\text{e}}$ series.
Since the map (\ref{def using projection}) depends only on 
the Bergman kernel, the expression	(\ref{BK for higher genera})
defines the quantization map (\ref{def using projection}) for 
the Riemann surfaces.

%The above Bergman kernel has a sort of the locality.
% which is needed for 
%(\ref{property of rectangular regularization}) to be satisfied. 
%To see this property, we evaluate
%the absolute value of the nonvanishing $(1,1)$ element of 
%$K^{(N)}(z,w)$ as
%\als
%{
%	|K^{(N)}_{11}(z,w)|
%	&\leq
%	\sum_{\gamma\in\Gamma}
%	\left|\frac{d\gamma(z)}{dz}\right|^{(1+N)/2}
%	\left|\frac{(1-\gamma(z)\bar w)^2}{(1-|z|^2)(1-|w|^2)}\right|^{-(N+1)/2}\\
%	&=
%	\sum_{\gamma\in\Gamma}
%	\left|\frac{(1-\gamma(z)\bar w)^2}{(1-|\gamma(z)|^2)(1-|w|^2)}\right|^{-(N+1)/2}\\
%	&=
%	\sum_{\gamma\in\Gamma}
%	\exp\left[-\frac{N+1}{2}\log
%	\left\{1+\frac{|\gamma(z)-w|^2}{(1-|\gamma(z)|^2)(1-|w|^2)}\right\}\right]\\
%	&=
%	\frac{2\pi}{N}(1-|w|^2)^2\sum_{\gamma\in\Gamma}
%	\delta^{(2)}(\gamma(z)-w)+O(N^{-2}).
%\label{bound for K}
%}
%Here, we used (\ref{relation for SU(1,1) action}) at the second equality.
%Note that $\gamma$ satisfying $\gamma(z)=w$ exists
%if and only if the two orbits of $\Gamma$ containing 
%$z$ and $w$ coincide with each other.
%Then, the delta function in the last line of (\ref{bound for K})
%behaves like the periodic delta function with respect to $\Gamma$
%and shows the locality of $K^{(N)}(z,w)$ on $\mathcal{M}$
%in the large-$N$ limit.
%By using the estimation (\ref{bound for K}), it is easy to see that 
%(\ref{property of rectangular regularization}) is satisfied for
%(\ref{def using projection}).

%%%%%%%%%%%%%%%%%%%%%%%%%%%%%%%%%%%%%%%%%%%%%%%%%%%%%%
\section{Conclusion and discussion}
In this paper, we considered the Berezin-Toeplitz quantization for
local sections of nontrivial complex line bundles on Riemann surfaces.
This corresponds to the matrix regularization of fields with 
nonvanishing $U(1)$ charges in a nontrivial gauge flux.
We argued that such fields are 
naturally mapped to rectangular matrices, while fields with 
vanishing charge are mapped to square matrices.
We also showed that these mappings are embedded in the 
Berezin-Toeplitz quantization in a $U(2)$ gauge theory in a unified way.
We then explicitly constructed those mappings for the sphere and the torus
and also discussed a possible extension to Riemann surfaces with 
higher genera. 
For the case of the sphere, we showed that 
this mapping reproduces the well-known fuzzy spherical harmonics.
For the case of the torus, we found that the mapping produces 
rectangular matrices written in terms of elliptic functions and 
we also proposed a matrix Laplacian, which reproduces 
the spectrum on the commutative torus in the large-$N$ limit.

In our examples of the sphere and the torus,
the operation $\circ$, which is defined in
(\ref{eq:4-2-83}) or
(\ref{angular momentum on rectangular matrix}),
was used in constructing the matrix Laplacians.
This operation gives another module structure of the matrix algebra 
satisfying
\begin{align}
&A \circ (BF) = [A, B]\circ F + B (A \circ F), \nonumber\\
&(AB)\circ F = A (B \circ F ) + (A \circ F ) B, \nonumber \\
&A\circ (B\circ F)-B\circ (A\circ F)= [A, B]\circ F,
\label{property of circ}
\end{align}
where $A$ and $B$ are square matrices and $F$ is a rectangular matrix.
If there exists a classical (commutative) counterpart of this operation,
it should satisfy the classical versions of these properties.
Unfortunately, we could not find such object and
our present optimal choices are
\begin{align}
&\{ f, a \}_1 := W^{\mu \nu } (\partial_\mu f) (D_{\nu}a),  \\
&\{ f, a\}_2 := W^{\mu \nu } (\partial_\mu f) (D_{\nu}a) -i Qfa,
\end{align}
where $f$ and $a$ are arbitrary smooth function and local section 
with charge $Q$, respectively, and $W^{\mu\nu}$ is a Poison tensor, 
given by the inverse of the symplectic form.
$\{  \; , \; \}_1$ only satisfies the first and the second properties of
(\ref{property of circ}), while $\{  \; , \; \}_2$ satisfies
the first and the third, provided that the 
quantization condition $[D_\mu, D_\nu]a = -i Q \omega_{\mu \nu}a$ is satisfied.
These operations cannot be the classical counterpart of $\circ$,
because the violations of the classical counterpart of (\ref{property of circ})
contradict with the properties (\ref{properties of MR}) and
(\ref{property of rectangular regularization}).
Probably, we would need another structure to construct the classical 
operation in general.
Nevertheless, we found that the continuum Laplacian on the sphere 
is proportional to $\{x_i , \{x_i, \;\; \}_2 \}_2$, while 
that on the torus is proportional to
$\{\bar{u} , \{u, \;\; \}_1 \}_1+\{\bar{v} , \{v, \;\; \}_1 \}_1$, 
and we consider that these brackets may still have meanings when 
some special functions are put in the first slots.
It should be important to understand this correspondence 
to find the description of the Laplacian in 
a general matrix geometry. 

The mapping for the local sections  we considered in this paper  
should also be relevant for describing D-branes with gauge fluxes.
In particular, it is known that the Berezin-Toeplitz quantization naturally 
appears in non-BPS D-brane systems with tachyon condensations
\cite{Asakawa:2001vm,Terashima:2005ic}
(see also \cite{Asakawa:2018gxf, Terashima:2018tyi}). 
In this context, introducing non-Abelian gauge groups seems to 
be quite natural,
so that our formulation using the $U(2)$ gauge theory will be 
naturally understood. This will be studied elsewhere.
It is important to understand properties of the quantization with 
a general gauge group in order to understand its implications in the 
D-brane systems.

Our results will also shed light on the problem of describing curved spaces
in the matrix models. For example, by using the mapping for rectangular matrices, 
it will be possible to generalize the work 
\cite{Ishiki:2006yr, Ishii:2008ib} and construct 
the large-$N$ reduction for nontrivial $U(1)$ bundles on Riemann surfaces,
called Seifert manifolds.

%%%%%%%%%%%%%%%%%%%%%%%%%%%%%%%%%%%%%%%%%%%%%%%%%%%%%%%%%%%%%
\section*{Acknowledgments}
%%%%%%%%%%%%%%%%%%%%%%%%%%%%%%%%%%%%%%%%%%%%%%%%%%%%%%%%%%%%%
We thank K.~Hasebe, Y.~Hatsugai, M.~Honda and N.~Ishibashi  for valuable discussions.
The work of G. I. was supported, in part, 
by Program to Disseminate Tenure Tracking System, 
MEXT, Japan and by KAKENHI (16K17679 and 19K03818).

%%%%%%%%%%%%%%%%%%%%%%%%%%%%%%%%%%%%%%%%%%%%%%%%%%%%
\begin{appendix}
\numberwithin{equation}{section}
\setcounter{equation}{0}

%%%%%%%%%%%%%%%%%%%%%%%%%%%%%%%%%%%%%%%%%%%%%%%%%%%%%%
\section{Dirac operator on Riemann surfaces with magnetic fluxes}
\label{Dirac operator on Riemann surface}
In this appendix, we construct a Dirac operator on a 
general Riemann surface $\mathcal{M}$ with a magnetic flux.

Let $z$ be a local complex coordinate on an open subset 
$U\subset\mathcal{M}$.
We define the standard complex structure $J$ on $\mathcal{M}$ by
$J(\partial_z)=i\partial_z$ and 
$J(\partial_{\bar z})=-i\partial_{\bar z}$.
Note that this definition does not depend on the choice of the local 
coordinate. Let $g$ be a K$\ddot{\text{a}}$hler metric on $\mathcal{M }$
compatible with $J$. On $U$, we have
$g_{zz}=g_{\bar z\bar z}=0$ and we can write $g$ as
\beq
	\label{metric on Riemann surface}
	g
	=
	2g_{z\bar z}dzd\bar z,
\eeq
where $g_{z \bar z}=g(\partial_z,\partial_{\bar z})$.
We define a symplectic form on $\mathcal{M}$ by 
$\omega(\,\cdot\,,\,\cdot\,)=g(\,J\cdot\,,\,\cdot\,)$.
In terms of the local coordinate,
we can write $\omega$ as
\beq
	\label{symplectic form on Riemann surface}
	\omega
	=
	ig_{z\bar z}dz\wedge d\bar z.
\eeq
We choose a $U(1)$ gauge field $A$ which satisfies $\omega=VdA$,
as explained in section 2.1.

Let $e_a$ be the zweibein for the K$\ddot{\text{a}}$hler metric
(\ref{metric on Riemann surface}). They are explicitly given by
\als
{
	&e_1
	=
	\frac{1}{\sqrt{2g_{z\bar z}}}(\partial_z+\partial_{\bar z}),\\
	&e_2
	=
	\frac{i}{\sqrt{2g_{z\bar z}}}(\partial_z-\partial_{\bar z}).
}
Note that from the positivity of the metric, $g_{z\bar z}$ is always 
positive.
%We define the gamma matrices on $U$ by
%\als
%{
%	&\gamma_z
%	=
%	\sqrt{\frac{g_{z\bar z}}{2}}(\sigma_1-i\sigma_2)
%	=
%	\sqrt{2g_{z\bar z}}\left(\begin{array}{cc}0&0\\1&0\end{array}\right),\\
%	&\gamma_{\bar z}
%	=
%	\sqrt{\frac{g_{z\bar z}}{2}}(\sigma_1+i\sigma_2)
%	=
%	\sqrt{2g_{z\bar z}}\left(\begin{array}{cc}0&1\\0&0\end{array}\right).
%}
The inverse $\theta_a$ of $e_a$ is given by
\als
{
	&\theta_1
	=
	\sqrt{\frac{g_{z\bar z}}{2}}(dz+d{\bar z}),\\
	&\theta_2
	=
	\frac{1}{i}\sqrt{\frac{g_{z\bar z}}{2}}(dz-d{\bar z}).
}
The spin connection is determined by
\beq
	\Omega^a{}_b\wedge\theta^b+d\theta^a=0,
\eeq
and $\Omega_{ab}=-\Omega_{ba}$. By solving this equation,
we find
\beq
	\label{spin connection on Riemann surface}
	\Omega_{12}
	=
	\frac{i}{2}(\partial_z\log g_{z\bar z}\,dz-\partial_{\bar z}\log g_{z\bar z}\,d\bar z).
\eeq

Any K$\ddot{\text{a}}$hler metric is locally written as
\beq
	\label{Kahler potential}
	g_{z\bar z}
	=
	\partial_z\partial_{\bar z}\rho,
\eeq
in terms of the K$\ddot{\text{a}}$hler potential $\rho$, which 
is a real function defined locally on $U$.
%However, for a given $g$, $\rho$ is not unique. 
The geometric structures we introduced above can also be 
expressed in terms of $\rho$.
For example, from (\ref{symplectic form on Riemann surface}) and
$\omega=VdA$, the gauge field $A$ is given by
\beq
	\label{U(1) connection on Riemann surface}
	A
	=
	-\frac{i}{2V}(\partial_z\rho\,dz-\partial_{\bar{z}} \rho\,d\bar z),
\eeq
up to the gauge transformation.

For $\mathcal{M}$, the Dirac operator $D$ defined by
(\ref{Dirac op}) flips the chirality, so that it has only off-diagonal elements.
Using the above data, we can express $D$ as
\als
{
	\label{general form of Dirac operator}
	&D^+=
	i\sqrt{\frac{2}{g_{z\bar z}}}
	\left\{\partial_{\bar z}+\frac{1}{2}\partial_{\bar z}
	\left(\frac{N}{V}\rho+\frac{1}{2}\log g_{z\bar z}\right)\right\},\\
	&D^-=
	i\sqrt{\frac{2}{g_{z\bar z}}}
	\left\{\partial_z-\frac{1}{2}\partial_z
	\left(\frac{N}{V}\rho-\frac{1}{2}\log g_{z\bar z}\right)\right\},
}
where $D^\pm$ are the matrix elements of $D$ acting 
on the spaces with chirality $\pm 1$, respectively.
%By solving the Dirac equation $D^\pm\psi^\pm=0$, we find that
%\beq
%	\label{Dirac zero mode on Riemann surface}
%	\psi^\pm
%	=
%	\exp\left[\mp\frac{1}{2}\left(\frac{N}{V}\rho
%	\pm\frac{1}{2}\text{log}\,\partial_z\partial_{\bar z}\rho\right)\right]h^\pm.
%\eeq
%Here, $h^+$ and $h^-$ are 
%holomorphic and anti-holomorphic sections of $L_N$.
%This is the general form of the Dirac zero modes.

%%%%%%%%%%%%%%%%%%%%%%%%%%%%%%%%%%%%%%%%%%%%%%%%%%%%
\section{Bergman kernel}
\label{Bergman kernel}

In this section, we give a formulation of the maps
(\ref{Toeplitz operator}) and (\ref{rectangular map})
in terms of the Bergman kernel.

Let us consider a product $\varphi^{(Q)}\psi^{(N)}_I$ of 
a charged scalar fields $\varphi^{(Q)}$ with charge $Q$ and a
Dirac zero mode $\psi^{(N)}_I$ with charge $N$, where 
$I=1,2, \cdots, N$.
This product has the total charge $N'=N+Q$, and can be 
expanded in terms of the Dirac eigen modes with charge $N'$ as
\begin{align}
	\varphi^{(Q)} \psi^{(N)}_I = \sum_{J=1}^{N'}c_{IJ} \psi^{(N')}_J + \cdots,
\end{align}
where $\cdots$ stands for the terms of non-zero modes.
If $\psi^{(N)}_I$ and $\psi^{(N')}_J$ are orthonormal basis of the 
zero modes, the coefficients $c_{IJ}$ can be extracted as
$c_{IJ} = (\psi^{(N')}_J, \varphi^{(Q)} \psi^{(N)}_I)$.
This is just the map (\ref{rectangular map}) for $Q\neq 0$ 
and (\ref{Toeplitz operator}) for $Q=0$.
Thus, those maps are obtained as actions of the charged scalar fields 
accompanied with the projections 
$\Pi^{(N')}$ and $\Pi^{(N)}$
onto the spaces of the zero modes \cite{Hawkins:1998nj}:
\begin{align}
	\hat T_{NN'}(\varphi^{(Q)}) = \Pi^{(N')} \varphi^{(Q)} \Pi^{(N)}.
	\label{def using projection}
\end{align}
The projections are given by the so-called Bergman kernel,
\begin{align}
	K^{(N)}(z, w) := \sum_{I=1}^N \psi_I^{(N)} (z) \psi_I^{(N)\dagger} (w).
	\label{def of Bergman kernel}
\end{align}
Here, the spinor indices are not contracted, so that 
$K^{(N)}(z, w)$ is a $2\times 2$ matrix with those indices.
The projection $\Pi^{(N)} $ is then defined by
\beq
	(\Pi^{(N)}\psi)(z)
	=
	\int_{\mathcal{M}}\omega(w)\,K^{(N)}(z,w)\psi(w).
	\label{projection for KerD}
\eeq
Note that the original expression (\ref{rectangular map}) is just the 
matrix representation of (\ref{def using projection})\footnote{
In our convention, the matrix representation of $\hat T_{NN'}$
is the transpose of $T_{NN'}$.}:
\als
{
	\label{definition of T op 2}
	T_{NN'}(\varphi^{(Q)})_{IJ}
	=
	(\psi^{(N')}_{J},\hat T_{NN'}(\varphi^{(Q)})\psi^{(N)}_{I})
	=
	\int_{\mathcal{M}}\omega\,\psi^{(N')\dagger}_{J}\cdot\varphi^{(Q)} \psi^{(N)}_{I}.
}

%%%%%%%%%%%%%%%%%%%%%%%%%%%%%%%%%%%%%%%%%%%%%%%%%%%%
\section{Proof of (\ref{property of rectangular regularization})}
\label{Proof of 1.3}
In this appendix, we give a proof of (\ref{property of rectangular regularization})
following \cite{Hawkins:1998nj}.

We first show that Dirac eigen modes with charge $N$ have 
a large energy gap in the large-$N$ limit.
Let $D$ be a Dirac operator for the charged spinors.
Let us consider the action of $D$ on a two-component spinor
$\chi$. Since $D$ generally flips the chirality, $D$ can be 
represented as
\begin{align}
D\chi =
\left(
\begin{array}{cc}
0 & D^- \\
D^+ & 0 \\
\end{array}
\right)
\left(
\begin{array}{c}
\chi^+ \\
\chi^- \\ 
\end{array}
\right),
\end{align}
where $\chi^+$ and $\chi^-$ are the positive and negative chirality modes of 
$\chi$, respectively.
Below, we assume that $D$ is normalized such that 
it is Hermitian and $(D^{\pm})^\dagger = D^\mp$. 
If $\chi$ is a normalized eigen mode of $D$ with eigenvalue $E$, we have
\begin{align}
D^+ \chi^+ = E\chi^-,  \;\;\; D^- \chi^- = E\chi^+,  
\;\;\; 
(\chi, \chi)=1.
\end{align}
Then, we can estimate the energy $E$ as
\begin{align}
E^2 &= E^2 (\chi, \chi) \nonumber\\
&= E^2 \int \omega \left( |\chi^+|^2 + |\chi^-|^2 \right) \nonumber\\
&=  \int \omega \left( |D^+\chi^+|^2 + |D^+\chi^-|^2 
+\bar\chi^{-} [D^+, D^-]\chi^- \right) \nonumber\\
&\geq \int \omega\, \bar\chi^{-} [D^+, D^-]\chi^-.
\label{estimating energy gap}
\end{align}
Now, we can assume that $[D^+, D^-]=-iF_{+-}+ {O}(N^0)$ 
is positive  in the large-$N$ limit without loss of generality, since
if it is negative, we can just exchange $\pm$ in the above calculation
and $[D^-, D^+]$ is positive there.
If $\chi$ is a nonzero mode, $\chi^-$ is nonvanishing.
In this case, the equation (\ref{estimating energy gap}) shows that
$E^2$ is bounded from below by a positive
quantity of order $N$. Thus, we found the energy gap,
\begin{align}
|E| \geq { O}(N^{1/2}).
\label{energy gap}
\end{align}

%We next introduce a projection operator onto the Dirac zero modes.
%Let $\{ \psi^{(N)}_{nI}\}$ be an orthonormal basis of the Dirac 
%eigen modes with charge $N$, where $n=0,1,\ldots,$ labels eigenvalues and 
%$I=1,2,\ldots,$ labels degeneracies for each energy level.
%The index $n$ shall be defined such that the corresponding 
%eigenvalues are ordered as $|E_0|\leq|E_1|\leq\cdots$.
%In particular, $\psi^{(N)}_{0I}$ shall be the zero modes and $E_0=0$.
%The index theorem implies that the degeneracy of the zero modes is $N$.

%Note that the original expression (\ref{rectangular map}) is just the 
%matrix representation of this:
%\als
%{
%	\label{definition of T op 2}
%	T_{NN'}(\varphi)_{IJ}
%	=
%	(\psi^{(N')}_{0J},\Pi^{(N')} \varphi \Pi^{(N)}\psi^{(N)}_{0I})
%	=
%	\int_{\mathcal{M}}\omega\,\psi^{(N')\dagger}_{0J}\varphi \psi^{(N)}_{0I}.
%}

Now, we prove (\ref{property of rectangular regularization}) for 
arbitrary smooth function $f$ and section $a$.
From (\ref{Toeplitz operator}), (\ref{rectangular map})
and (\ref{projection for KerD}), we have
\als
{
	[T_N(f)T_{NN'}(a)-T_{NN'}(f\cdot a)]_{IJ}
	=
	\int_{\mathcal{M}}\omega\,\psi^{(N')\dagger}_J
	\cdot a[\Pi^{(N)},f]\psi^{(N)}_I.
}
Thus,
\als
{
	\label{condition 1}
	|[T_N(f)T_{NN'}(a)-T_{NN'}(f\cdot a)]_{IJ}|
	&\leq
	\int_{\mathcal{M}}\omega\,|\psi^{(N')\dagger}_J
	\cdot a[\Pi^{(N)},f]\psi^{(N)}_I| \\
	&\leq
	\|a\|\,\|[\Pi^{(N)},f]\|
	\int_{\mathcal{M}}\omega\,|\psi^{(N')\dagger}_J \cdot\psi^{(N)}_I|.
}
Here, we have defined the norms for local sections and operators on
spinors by
\als
{
	&\|a\|
	=\sup_{x\in\mathcal{M}}|a(x)|,\\
	&\|[\Pi^{(N)},f]\|
	=\sup_{\|\psi\|=1}\|[\Pi^{(N)},f]\psi\|.
}
From the orthonormality
$\int_{\mathcal{M}}\omega\,\psi^{(N)\dagger}_I \cdot\psi^{(N)}_J=\delta_{IJ}$,
the last integral in (\ref{condition 1}) is of $O(N^0)$.
Also, the norm $\|a\|$ is finite in the large-$N$ limit.
The only nontrivial factor is $\|[\Pi^{(N)},f]\|$ and we will 
estimate this in the following.
Let us consider an operator $(1+\alpha D^2)^{-1}$,
where $\alpha$ is an $N$-independent positive number.
Note that $1+\alpha E^2>0$ for any eigenvalue $E$,
so that the operator $(1+\alpha D^2)^{-1}$ is well-defined.
Since $E\geq O(N^{1/2})$ except for the case $E=0$ as shown in
(\ref{energy gap}),
%\als
%{
%	\label{projection}
%	(1+\alpha D^2)^{-1}\psi^{(N)}_{nI}
%	=
%	\delta_{n0}\psi^{(N)}_{0I}+\alpha^{-1}O(N^{-1}).
%}
the operator $(1+\alpha D^2)^{-1}$ behaves as
a projection onto $\text{Ker}D$ for sufficiently large values of $N$.
Hence, we obtain
\beq
	\Pi^{(N)}
	=
	(1+\alpha D^2)^{-1}+\alpha^{-1}O(N^{-1}).
\eeq
Thus, the problem of estimating $[\Pi^{(N)},f]$ reduces to 
that of $[(1+\alpha D^2)^{-1},f]$.
In order to evaluate the latter, we first rewrite this as
\als
{
	[(1+\alpha D^2)^{-1},f]
	&=
	(1+\alpha D^2)^{-1}[f,(1+\alpha D^2)](1+\alpha D^2)^{-1} \\
	&=
	\alpha(1+\alpha D^2)^{-1}[f,D^2](1+\alpha D^2)^{-1}.
}
By using the Leibniz rule 
$D(f\psi)=(i\sigma^\mu\partial_\mu f)\psi+f(D\psi)$,
where $\sigma^\mu=\sigma^a\theta^\mu_a$,
we also have
\als
{
	[f,D^2]
	&=
	fD^2-D((i\sigma^\mu\partial_\mu f)+fD) \\
	&=
	-D(i\sigma^\mu\partial_\mu f)-(i\sigma^\mu\partial_\mu f)D \\
	&=
	-\{D,(i\sigma^\mu\partial_\mu f)\}.
}
Thus, we obtain
\beq
	\label{commutator}
	[\Pi^{(N)},f]
	=
	-\alpha(1+\alpha D^2)^{-1}\{D,(i\sigma^\mu\partial_\mu f)\}(1+\alpha D^2)^{-1}
	+\alpha^{-1}O(N^{-1}).
\eeq
This gives the following estimation:
\als
{
	\|[\Pi^{(N)},f]\|
	&\leq
	2\alpha\|(1+\alpha D^2)^{-1}D\|\,\|(1+\alpha D^2)^{-1}\|
	\,\|\sigma^\mu\partial_\mu f\| \\
	&\leq
	2|E_1|^{-1}\,\|\sigma^\mu\partial_\mu f\|
	\label{estimation of pi f}
}
where $E_1$ is the smallest nonzero eigenvalue of $D$.
The last inequality is obtained as follows. 
For any eigenvalue $E$, we have the relation,
\als
{
	\frac{|E_1|}{1+\alpha E^2_1}-\frac{|E|}{1+\alpha E^2}
	=
%	\frac{|E_1|\{1+\alpha (E_n)^2\}-|E_n|\{1+\alpha (E_1)^2\}}
%	{\{1+\alpha (E_1)^2\}\{1+\alpha (E_n)^2\}}\\
%	&=
	\frac{(|E|-|E_1|)(\alpha|E_1||E|-1)}{\{1+\alpha E^2_1\}\{1+\alpha E^2\}}	
	\geq
	0.
}
This implies that
\beq
	\|(1+\alpha D^2)^{-1}D\|
	\leq
	\frac{|E_1|}{1+\alpha E^2_1}
	\leq
	\alpha^{-1}|E_1|^{-1},
\eeq
which, together with the obvious relation $\|(1+\alpha D^2)^{-1}\|\leq1$, 
leads to the second inequality in (\ref{estimation of pi f}).
By applying (\ref{estimation of pi f}) to (\ref{condition 1}), we finally obtain
\beq
	|[T_N(f)T_{NN'}(a)-T_{NN'}(f\cdot a)]_{IJ}|
	\leq
	2|E_1|^{-1}\|a\|\,\|\sigma^\mu\partial_\mu f\|
	\int_{\mathcal{M}}\omega\,|\psi^{(N')\dagger}_J \cdot\psi^{(N)}_I|.
\eeq
Since $|E_1|\geq O(N^{1/2})$, the right-hand side vanishes in the 
large-$N$ limit and we find that
(\ref{property of rectangular regularization}) is indeed satisfied.

%%%%%%%%%%%%%%%%%%%%%%%%%%%%%%%%%%%%%%%%%%%%%%%%
\section{Monopole harmonics}
\label{appendix for monopole harmonics}

In this appendix, we review the definition of the monopole harmonics.
See \cite{Wu:1976ge, Wu:1977qk} for more details.

We first introduce linear operators which is locally defined on $U_1$ and $U_2$ as
\als
{
	\label{angular momentum on local section}
	&\mathcal{L}^{(Q)}_1
	=i(\sin\phi\,\partial_\theta+\cot\theta\cos\phi\,\partial_\phi)
	-\frac{Q}{2}\frac{1\mp\cos\theta}{\sin\theta}\cos\phi,\\
	&\mathcal{L}^{(Q)}_2
	=i(-\cos\phi\,\partial_\theta+\cot\theta\sin\phi\,\partial_\phi)
	-\frac{Q}{2}\frac{1\mp\cos\theta}{\sin\theta}\sin\phi,\\
	&\mathcal{L}^{(Q)}_3
	=-i\partial_\phi\mp \frac{Q}{2},
}
where the upper and lower signs represent the expressions
on $U_1$ and $U_2$, respectively.
These operators are the angular momentum operator in the presence of
a monopole with magnetic charge $Q/2$ at the origin and 
reduces to the ordinary angular momentum operators when $Q=0$.
In fact, $\mathcal{L}^{(Q)}_A$ $(A=1,2,3)$ satisfy the $SU(2)$ algebra
$[\mathcal{L}^{(Q)}_A, \mathcal{L}^{(Q)}_B]=i\epsilon_{ABC}L^{(Q)C}$
and give a representation of the Lie algebra of $SU(2)$ on
the space of charged scalar fields which transform as (\ref{gauge tra of section}).
In particular, the action of $\mathcal{L}^{(Q)}_A$ is covariant under
the gauge transformation (\ref{gauge tra of section}).

The monopole harmonics
$Y^{(Q)}_{lm}$ $(l=|Q|/2,|Q|/2+1,\ldots,\infty, m=-l,-l+1,\ldots,l)$
are defined as the standard basis of this representation space
which satisfies
\als
{
	\label{definition1 of Y}
	&(\mathcal{L}^{(Q)}_A)^2Y^{(Q)}_{lm}=l(l+1)Y^{(Q)}_{lm},\\
	&\mathcal{L}^{(Q)}_3 Y^{(Q)}_{lm}=mY^{(Q)}_{lm},
}
and the orthonormal condition
\beq
	\int_{S^2}\omega\, (Y^{(Q)}_{lm})^\ast Y^{(Q)}_{l'm'}
	=\delta_{ll'}\delta_{mm'},
\eeq
for a fixed $l$, where $\omega$ is the volume form for
the metric (\ref{metric and volume form on S^2}).
The concrete expression of $Y^{(Q)}_{lm}$ is
\beq
	\label{monopole harmonics}
	Y^{(Q)}_{lm}
	=(-1)^{l-m-Q}\sqrt{\frac{2l+1}{2\pi}}
	\mel*{l-m}{e^{-i\theta L^{(l)}_2}}{l\frac{Q}{2}}e^{i(\pm Q/2+m)\phi},
\eeq
where $L^{(l)}_A$ are the $(2l+1)$-dimensional irreducible representation of
the generators
of $SU(2)$ and $\ket*{lm}$ are the standard basis of the representation space.
Again, the upper and lower signs represent the expressions defined
on $U_1$ and $U_2$, respectively.

%At the south pole $\theta=\pi$, $Y^{(Q)}_{1lm}$
%which is defined on $U_1$ is not zero only for $m=Q/2$
%and is proportional to a factor $e^{iQ\phi}$.
%Since $\phi$ varies in the range $0<\phi\leq 2\pi$,
%$Y^{(Q)}_{1lm}$ is not single-valued at the south pole
%and therefore is well-defined only on $U_1$.
%Similarly, at the north pole $\theta=0$,
%$Y^{(Q)}_{2lm}$ which is defined on $U_2$
%is not zero only for $m=-Q/2$ and is proportional to a factor $e^{-iQ\phi}$.
%Therefore, $Y^{(Q)}_{2lm}$ is well-defined only on $U_2$.

The following formula is very useful.
\beq
	\label{three monopole harmonics}
	\int_{S^2}\omega\,
	(Y^{(Q_1)}_{l_1m_1})^\ast Y^{(Q_2)}_{l_2m_2}Y^{(Q_3)}_{l_3m_3}
	=\sqrt{\frac{(2l_2+1)(2l_3+1)}{2\pi(2l_1+1)}}
	C^{l_1m_1}_{l_2m_2l_3m_3}C^{l_1Q_1/2}_{l_2Q_2/2l_3Q_3/2},
\eeq
where $C^{l_1m_1}_{l_2m_2l_3m_3}$ is the Clebsch-Gordan coefficient.
For the gauge invariance of the left hand side,
$Q_1=Q_2+Q_3$ must hold.

%%%%%%%%%%%%%%%%%%%%%%%%%%%%%%%%%%%%%%%%%%%%%%%%
\section{Fuzzy spherical harmonics}
\label{appendix for fuzzy spherical harmonics}

In this appendix, we review the definition of the fuzzy spherical harmonics
\cite{Grosse:1995jt,Baez:1998he,Dasgupta:2002hx}.
See \cite{Ishiki:2006yr, Ishii:2008ib} for more details.

We first define linear operators on $M_{N\times N'}(\mathbf{C})$ by
\beq
	\label{angular momentum on rectangular matrix}
	L^{(JJ')}_A\circ M:=L^{(J)}_AM-ML^{(J')}_A.
\eeq
for $M\in M_{N\times N'}(\mathbf{C})$,
where $L^{(J)}_A$ are the $(2J+1)$-dimensional representation of
the $SU(2)$ generators.
Then, $L^{(JJ')}_A\circ$ satisfy the $SU(2)$ algebra
$[L^{(JJ')}_A\circ,L^{(JJ')}_B\circ]=i\epsilon_{ABC}L^{(JJ')C}\circ$
and therefore give the $(N\times N')$-dimensional representation of
the Lie algebra of $SU(2)$.

The fuzzy spherical harmonics 
$\hat Y_{lm(JJ')}$ $(l=|J-J'|,|J-J'|+1,\ldots,J+J', m=-l,-l+1,\ldots,l)$
are defined as the standard basis of this representation space which
satisfies
\als
{
	\label{definition1 of hat Y}
	&(L^{(JJ')}_A\circ)^2\hat Y_{lm(JJ')}=l(l+1)\hat Y_{lm(JJ')},\\
	&L^{(JJ')}_3\circ \hat Y_{lm(JJ')}=m\hat Y_{lm(JJ')},
}
and the orthonormal condition
\beq
	\label{definition2 of hat Y}
	\frac{1}{N}\tr \{(\hat Y_{lm(JJ')})^\dagger \hat Y_{l'm'(JJ')}\}
	=\delta_{ll'}\delta_{mm'},
\eeq
for a fixed $l$. Here, the trace is defined over $N\times N'$ matrices.
In terms of the basis $\{\op*{Jr}{J'r'}\}$, they are expressed as
\beq
	\hat Y_{lm(JJ')}
	=\sqrt{N}\sum^J_{r=-J}\sum^{J'}_{r'=-J'}(-1)^{-J+r'}C^{lm}_{JrJ'-r'}\op{Jr}{J'r'}.
\eeq

%%%%%%%%%%%%%%%%%%%%%%%%%%%%%%%%%%%%%%%%%%%%%%%%%%%%%
\section{Detailed calculation of the normalization factor $\mathcal{N}_r$}
\label{normalization calculation}
In this appendix, we derive \eqref{eq:4-2-45}.

From of the explicit form of the Jacobi-theta function, we first write
\begin{equation}
\begin{aligned}
(\psi^{(N)+}_{r'}, \psi^{(N)+}_r) = 2 \Im (\tau) \mathcal{N}_r \mathcal{N}_{r'} \sum_{l,l' \in \mathbf{Z}} & \mathrm{e}^{iN\pi \left\{ \tau (l + r/N)^2 - \bar{\tau}(l' + r'/N)^2 \right\}} \int_0^1 dx \int_0^1 dy  \\
& \times \mathrm{e}^{-2N \pi \left\{ \Im(z+\zeta) \right\}^2 /\Im \tau} \mathrm{e}^{i2N\pi \left\{ (l + r/N)(z + \zeta) - (l' + r'/N)(\bar{z} + \bar{\zeta}) \right\}}.
\end{aligned}
\label{eq:A-1-1}
\end{equation}
Then, by shifting the integration variable as $z \to z - \zeta$
and substituting $z = x + \tau y$, we obtain
\begin{equation}
\begin{aligned}
(\psi^{(N)+}_{r'}, \psi^{(N)+}_r) = 2 \Im (\tau) \mathcal{N}_r \mathcal{N}_{r'} \sum_{l,l' \in \mathbf{Z}} & \mathrm{e}^{iN\pi \left\{ \tau (l + r/N)^2 - \bar{\tau}(l' + r'/N)^2 \right\}} \int_{\zeta_1}^{1 + \zeta_1} dx \int_{\zeta_2}^{1 + \zeta_2} dy  \\
& \times \mathrm{e}^{-2N \pi \Im (\tau) y^2} \mathrm{e}^{i2\pi \left\{ (r + Nl)\tau - (r' + Nl')\bar{\tau} \right\}y} \mathrm{e}^{i2\pi \left\{ r-r' + N(l-l') \right\}x}.
\end{aligned}
\label{eq:A-1-3}
\end{equation}
The integration over $x$ just produces 
the Kronecker delta factor $\delta_{r, r'} \delta_{l, l'}$.
Thus, by taking the summation over $l'$ and we obtain
\begin{equation}
(\psi^{(N)+}_{r'}, \psi^{(N)+}_r) = 2 \Im (\tau) \mathcal{N}_r^2 \delta_{r, r'} \sum_{l \in \mathbf{Z}} \mathrm{e}^{-2N\pi (l + r/N)^2 \Im\tau} \int_{\zeta_2}^{1 + \zeta_2} dy \, \mathrm{e}^{-2N \pi \Im (\tau) y^2} \mathrm{e}^{-4 \pi (r + Nl) \Im (\tau) y}. \label{eq:A-1-4}
\end{equation}
This can also be written in a compact form as 
\begin{equation}
(\psi^{(N)+}_{r'}, \psi^{(N)+}_r) = 2 \Im (\tau) \mathcal{N}_r^2 \delta_{r, r'} \sum_{l \in \mathbf{Z}} \int_{\zeta_2}^{1 + \zeta_2} dy \, \mathrm{e}^{-2N \pi \Im (\tau) \left(y+l+r/N\right)^2}. \label{eq:A-1-5}
\end{equation}
By again shifting the integration variable as $y \to y-l-r/N$,
 we obtain
\begin{equation}
(\psi^{(N)+}_{r'}, \psi^{(N)+}_r) = 2 \Im (\tau) \mathcal{N}_r^2 \delta_{r, r'} \sum_{l \in \mathbf{Z}} \int_{l +\frac{r}{N}+ \zeta_2}^{l +\frac{r}{N} + 1 + \zeta_2} dy \, \mathrm{e}^{-2N \pi \Im (\tau) y^2}. \label{eq:A-1-6}
\end{equation}
Since the $l$-dependence appears only in the integration range,
summing up all $l \in \mathbf{Z}$ is equivalent to extending 
the integration range to $(-\infty, \infty)$.
Thus, we finally arrive at a simple Gaussian integral. The final result is
\begin{equation}
(\psi^{(N)+}_{r'}, \psi^{(N)+}_r) = 2 \Im (\tau) \mathcal{N}_r^2 \delta_{r, r'} \int_{-\infty}^{\infty} dy \, \mathrm{e}^{-2N \pi \Im (\tau) y^2} = \sqrt{2 \Im \tau/N} \delta_{r,r'}.
\label{eq:A-1-6}
\end{equation}

%%%%%%%%%%%%%%%%%%%%%%%%%%%%%%%%%%%%%%%%%%%%%%%%%%%%%%%%%
\section{Orthonormal basis of local sections on the torus}
\label{Orthonormal basis of local sections on the torus}

In this appendix, we construct an orthonormal basis of 
local sections of the nontrivial line bundle with charge $Q$.

As the orthonormal basis, we consider a set of eigenfunctions of the Laplacian.
Let us consider the Laplacian for charge $Q \in \mathbf{N}$ given by
\begin{equation}
\begin{aligned}
\Delta := - 2g^{ab} D_a D_b 
= -2 \left( D_z D_{\bar{z}} + D_{\bar{z}} D_z \right),
\end{aligned}\label{eq:4-2-57}
\end{equation}
where
\begin{equation}
\begin{aligned}
D_z = \partial_z - \frac{ Q \pi}{2 \Im\tau} (\bar{z} + \bar{\zeta}), \\
D_{\bar{z}} = \partial_{\bar{z}} + \frac{ Q \pi}{2\Im\tau} (z + \zeta).
\end{aligned}\label{eq:4-2-58}
\end{equation}
%The commutator $[D_z, K_{\bar{z}}]$ gives a constant 
%field strength, so that these operators obey the Heisenberg algebra.
In the following, we will solve the eigenvalue problem,
\begin{equation}
\Delta \varphi^{(Q)}_n = E_n \varphi^{(Q)}_n \quad (n \in \mathbf{Z}_{\ge 0}),
\label{eq:4-2-59}
\end{equation}
to find the orthonormal eigen modes $\varphi^{(Q)}_n$ as well as 
the eigenvalues $E_n$.
Here, the eigen modes shall be ordered as
$E_n < E_{n+1} (\forall n \in \mathbf{Z}_{\ge 0})$.

We first introduce the creation-annihilation operators as
\als
{
	&\hat{a} := - i \sqrt{\frac{\Im\tau}{Q\pi}} D_{\bar{z}},\\
	&\hat{a}^{\dagger} = - i 	\sqrt{\frac{\Im\tau}{Q\pi}} D_{z},
	\label{eq:4-2-60}
}
which satisfy the commutation relation
\begin{equation}
\left[ \hat{a}, \hat{a}^{\dagger} \right] = 1. \label{eq:4-2-61}
\end{equation}
Then, \eqref{eq:4-2-59} can be expressed in terms of the number 
operator $\hat{N} := \hat{a}^{\dagger} \hat{a}$ as
\begin{equation}
\frac{4 Q \pi}{\Im\tau} \left( \hat{N} + \frac{1}{2} \right) \varphi^{(Q)}_n = E_n \varphi^{(Q)}_n .\label{eq:4-2-62}
\end{equation}
This is completely the same as the system of the 
1-dimensional harmonic oscillator.
Hence, from the standard argument, we find that
the normalized eigenfunctions and the eigenvalues are given by
\begin{equation}
\varphi^{(Q)}_n = \frac{\left( \hat{a}^{\dagger} \right)^n}{\sqrt{n!}} \varphi^{(Q)}_0 \quad \text{and} \quad E_n = \frac{4 Q \pi}{\Im\tau} \left( n + \frac{1}{2} \right) \quad (n \in \mathbf{Z}_{\ge 0}),
\label{eq:4-2-63}
\end{equation}
and the ground state is determined by
\begin{equation}
\hat{a} \varphi^{(Q)}_0 = 0 \quad \Leftrightarrow \quad D_{\bar{z}} \varphi^{(Q)}_0 = 0.
\label{eq:4-2-64}
\end{equation}
%Note that (\ref{eq:4-2-64}) is exactly the same equation as the Dirac zero mode %equation for the positive chirality mode.
Taking the boundary condition \eqref{eq:4-2-55} into account, 
we find the following form for the ground states:
\begin{equation}
\varphi^{(Q)}_{0,r} = \left( \frac{Q} {2 \Im\tau} \right)^{1/4} \mathrm{e}^{i Q \pi (z + \zeta) \frac{\Im(z + \zeta)}{\Im\tau}} \, \vartheta
	\left[
	\begin{array}{c}
	\frac{r}{Q} \\
	0
	\end{array}
	\right]
\left( Q(z + \zeta), \, Q \tau \right)
\label{eq:4-2-65}
\end{equation}
Here, the index $r=0,1,\cdots,Q-1$ labels the degeneracy of the ground states.
The normalization factors were determined by the orthonormality 
of $\varphi^{(Q)}_{0,r}$ with respect to the standard norm given by 
the integration with the symlectic form (\ref{eq:4-2-9}).
We can also calculate the excited modes from \eqref{eq:4-2-63}. 
The result is given by
\begin{equation}
\begin{aligned}
\varphi^{(Q)}_{n,r} = \frac{1}{\sqrt{2^n n!}} \left( \frac{Q} {2 \Im\tau} \right)^{1/4} \mathrm{e}^{i Q \pi (z + \zeta) \frac{\Im(z + \zeta)}{\Im\tau}} \sum_{l \in \mathbf{Z}} H_n \left( \sqrt{2Q \pi \Im\tau}\left( \frac{\Im(z + \zeta)}{\Im\tau} + l + \frac{r}{Q} \right) \right) \\
\times \mathrm{e}^{i\frac{\pi}{Q} (r + Ql)^2 \tau} \mathrm{e}^{i2 \pi (r + Ql)(z + \zeta)}.
\label{eq:4-2-66}
\end{aligned}
\end{equation}
where $H_n(x)$ is the Hermite polynomial 
satisfying $H_{n+1} (x) = 2xH_n (x) - H'_n(x)$.

The set 
$\{\varphi_{n,r}^{(Q)} | r= 0,1, \cdots, Q-1, \; n \in \mathbf{Z}_{\ge 0} \} $
forms an orthonormal basis of local sections with 
the twisted boundary condition (\ref{eq:4-2-55}).
The orthonormality of this basis is expressed as
\begin{align}
(\varphi_{n,r}^{(Q)}, \varphi_{n',r'}^{(Q)}) := \int \omega\,
\bar{\varphi}_{n,r}^{(Q)} \varphi_{n',r'}^{(Q)} = \delta_{nn'}\delta_{rr'}.
\label{orthonormality of local section basis on T2}
\end{align}

%%%%%%%%%%%%%%%%%%%%%%%%%%%%%%%%%%%%%%%%%%%%%%%%%%%%%%%
\section{Useful relations for the eigenstates of the Laplacian}
\label{Product of higher modes of Laplacian eigenstates}
In this appendix, 
we show some useful identities for the eigenstates of the Laplacian.

Let us consider the product
\begin{equation}
\varphi^{(Q)}_{n,s} (y, \bar{y}) \varphi^{(N)}_{m.r} (z, \bar{z}) = \frac{\left( \hat{a}^{\dagger}_y \right)^n}{\sqrt{n!}} \varphi^{(Q)}_{0,s} (y, \bar{y}) \frac{\left( \hat{a}^{\dagger}_z \right)^m}{\sqrt{m!}} \varphi^{(N)}_{0,r} (z, \bar{z})
\label{eq:A-2-1}
\end{equation}
of the eigenstates (\ref{eq:4-2-66}), 
where  $\hat{a}^{\dagger}_y$ and $\hat{a}^{\dagger}_z$ stand 
for the creation operators (\ref{eq:4-2-60}) acting on the complex variables $y$ and $z$, respectively\footnote{In this appendix, we use $y$ as a complex variable
exceptionally, while it is used as a real variable in the other sections.}.
By using the identity of the Jacobi-theta function,
%defined in \eqref{eq:4-2-40}.
\begin{equation}
\begin{aligned}
\vartheta
	\left[
	\begin{array}{c}
	\frac{s}{Q} \\
	0
	\end{array}
	\right]
\left(Q z_1, \, Q \tau \right)
\, \vartheta
	\left[
	\begin{array}{c}
	\frac{r}{N} \\
	0
	\end{array}
	\right]
\left(N z_2, \, N \tau \right)
&= \sum_{t=1}^{Q+N}
\vartheta
	\left[
	\begin{array}{c}
	\frac{r+s+Qt}{Q+N} \\
	0
	\end{array}
	\right]
\left(Q z_1 +N z_2, \, (Q+N) \tau \right)\\
&\qquad\times \,
\vartheta
	\left[
	\begin{array}{c}
	\frac{Ns-Qr+QNt}{QN(Q+N)} \\
	0
	\end{array}
	\right]
\left(QN(z_1 - z_2), \, QN(Q+N) \tau \right), \\
\end{aligned}
\label{eq:A-2-2}
\end{equation}
We rewrite (\ref{eq:A-2-1}) into
\begin{equation}
\varphi^{(Q)}_{0,s} (y, \bar{y}) \varphi^{(N)}_{0,r} (z, \bar{z}) = \frac{1}{\sqrt{N'}} \sum_{t=1}^{N'} \varphi^{(N')}_{0.r+s+Qt} (X,\bar{X}) \varphi^{(QNN')}_{0,Ns-Qr+QNt} (Y, \bar{Y}),
\label{eq:A-2-3}
\end{equation}
where we used $N'=N+Q$ and $X$ and $Y$ are defined by
\als
{
	&X := \frac{Qy+Nz}{N'},\\
	&Y :=\frac{y-z}{N'} - \zeta .
	\label{eq:A-2-4}
}
If we regard (\ref{eq:A-2-4}) as a change of variables from 
$(y, z)$ to $(X, Y)$, we can also convert 
 $\hat{a}^{\dagger}_y$ and $\hat{a}^{\dagger}_z$ to those 
in the $X$- and $Y$- coordinates:
\als
{
	&\hat{a}^{\dagger}_y
	= \sqrt{\frac{Q}{N'}} \hat{a}^{\dagger}_X + \sqrt{\frac{N}{N'}} \hat{a}^{\dagger}_Y,\\
	&\hat{a}^{\dagger}_z
	= \sqrt{\frac{N}{N'}} \hat{a}^{\dagger}_X - \sqrt{\frac{Q}{N'}} \hat{a}^{\dagger}_Y .
	\label{eq:A-2-6}
}
By using \eqref{eq:A-2-3} and \eqref{eq:A-2-6}, 
we calculate \eqref{eq:A-2-1} as
\begin{equation}
\begin{aligned}
&\varphi^{(Q)}_{n,s} (y, \bar{y}) \varphi^{(N)}_{m,r} (z, \bar{z})\\
&\quad= \frac{1}{\sqrt{n!m! N'}} \sum_{k=0}^n \sum_{l=0}^m \sum_{t=1}^{N'} (-1)^{m-l} \left(
	\begin{array}{c}
	n \\
	k
	\end{array}
	\right)
	\left(
	\begin{array}{c}
	m \\
	l
	\end{array}
	\right) \left( \frac{Q}{N'} \right)^{k/2} \left( \frac{N}{N'} \right)^{(n-k)/2} \left( \frac{N}{N'} \right)^{l/2} \left( \frac{Q}{N'} \right)^{(m-l)/2}
	\\ &\quad\qquad\qquad\times
	\left( \hat{a}^{\dagger}_X \right)^{k+l} \varphi^{(N')}_{0,r+s+Qt} (X,\bar{X}) \left( \hat{a}^{\dagger}_Y \right)^{n+m-k-l} \varphi^{(QNN')}_{0,Ns-Qr+QNt} (Y, \bar{Y}) \\
&\quad= \sum_{k=0}^n \sum_{l=0}^m (-1)^{m-l} \sqrt{\frac{(k+l)! (n+m-k-l)!}{n!m! N'}} \left(
	\begin{array}{c}
	n \\
	k
	\end{array}
	\right)
	\left(
	\begin{array}{c}
	m \\
	l
	\end{array}
	\right) \left( \frac{Q}{N'} \right)^{(k+m-l)/2} \left( \frac{N}{N'} \right)^{(l+n-k)/2}
	\\ &\quad\qquad\qquad\times
	\sum_{t=1}^{N'} \varphi^{(N')}_{k+l,r+s+Qt} (X,\bar{X}) \varphi^{(QNN')}_{n+m-k-l,Ns-Qr+QNt} (Y, \bar{Y}).
\end{aligned} \label{eq:A-2-7}
\end{equation}
Finally, we put $y=z$ and obtain
\begin{equation}
\begin{aligned}
&\varphi^{(Q)}_{n,s} (z, \bar{z}) \varphi^{(N)}_{m,r} (z, \bar{z})\\
&\quad= \sum_{k=0}^n \sum_{l=0}^m (-1)^{m-l} \sqrt{\frac{(k+l)! (n+m-k-l)!}{n!m! N'}} \left(
	\begin{array}{c}
	n \\
	k
	\end{array}
	\right)
	\left(
	\begin{array}{c}
	m \\
	l
	\end{array}
	\right) \left( \frac{Q}{N'} \right)^{(k+m-l)/2} \left( \frac{N}{N'} \right)^{(l+n-k)/2}
	\\ &\quad\qquad\qquad\times
	\sum_{t=1}^{N'} \varphi^{(N')}_{k+l,r+s+Qt} (z,\bar{z}) \varphi^{(QNN')}_{n+m-k-l,Ns-Qr+QNt} (-\zeta, -\bar{\zeta}).
\end{aligned} \label{eq:A-2-8}
\end{equation}

%%%%%%%%%%%%%%%%%%%%%%%%%%%%%%%%%%%%%%%%%%%%%%%%%%%%%
\section{Derivation of \eqref{eq:4-2-84}}
\label{Derivation in appendix}
In this appendix, we derive the equation \eqref{eq:4-2-84}.

For this purpose, we first need to compute  
\begin{equation}
A \circ T_{NN'}(\varphi^{(Q)}_{n,s})_{rr'} = \left( \varphi^{(N')}_{0,r'}, \varphi^{(Q)}_{n,s} \left[ A^{(N)}_{r \tilde{r}} \varphi^{(N)}_{0,\tilde{r}} \right] \right) - \left( \left[ \bar{A}^{(N')}_{\tilde{r}r'} \varphi^{(N')}_{0,\tilde{r}} \right], \varphi^{(Q)}_{n,s} \varphi^{(N)}_{0,r} \right) 
\label{eq:A-3-1}
\end{equation}
for $A=U, V$, where the inner product is defined in
(\ref{orthonormality of local section basis on T2}).
For $A=V$, we can calculate this as
\begin{equation}
\begin{aligned}
V^{(N)}_{r\tilde{r}} \varphi^{(N)}_{0,\tilde{r}} (x,y) &= \mathrm{e}^{-\frac{\pi}{2N}} \mathrm{e}^{-i2\pi\frac{r}{N}} (N/2)^{1/4} \mathrm{e}^{iN\pi y(x+iy)} \sum_{l\in \mathbf{Z}} \mathrm{e}^{-\frac{\pi}{N}(r+Nl)^2} \mathrm{e}^{i2\pi(r+Nl)(x+iy)} \\
&= \mathrm{e}^{-\frac{\pi}{2N}} (N/2)^{1/4} \mathrm{e}^{iN\pi y\left(x-\frac{1}{N}+iy+\frac{1}{N} \right)} \sum_{l\in \mathbf{Z}} \mathrm{e}^{-\frac{\pi}{N}(r+Nl)^2} \mathrm{e}^{i2\pi(r+Nl)\left(x-\frac{1}{N}+iy \right)} \\
&= \mathrm{e}^{-\frac{\pi}{2N}} \mathrm{e}^{i\pi y} \varphi^{(N)}_{0,r} (x-\frac{1}{N},y).
\end{aligned}
\label{eq:A-3-2}
\end{equation}
Similarly, we can obtain 
\begin{equation}
\bar{V}^{(N)}_{\tilde{r}r} \varphi^{(N)}_{0,\tilde{r}} (x,y) = \mathrm{e}^{-\frac{\pi}{2N}} \mathrm{e}^{-i\pi y} \varphi^{(N)}_{0,r} (x+\frac{1}{N},y).
\label{eq:A-3-3}
\end{equation}
By repeating a similar computation,  we obtain for $A=U$,
\begin{equation}
\begin{aligned}
U^{(N)}_{r\tilde{r}} \varphi^{(N)}_{0,\tilde{r}} (x,y)
&= \mathrm{e}^{-\frac{\pi}{2N}} \varphi^{(N)}_{0,r+1} (x,y)\\
&= \mathrm{e}^{-\frac{\pi}{2N}} (N/2)^{1/4} \mathrm{e}^{iN\pi y\left(x+iy \right)} \sum_{l\in \mathbf{Z}} \mathrm{e}^{-\frac{\pi}{N}(r+1+Nl)^2} \mathrm{e}^{i2\pi(r+1+Nl)\left(x+iy \right)} \\
&= \mathrm{e}^{-\frac{\pi}{2N}} (N/2)^{1/4} \mathrm{e}^{iN\pi y\left(x+iy \right)} \sum_{l\in \mathbf{Z}} \mathrm{e}^{-\frac{\pi}{N}(r+Nl)^2} \mathrm{e}^{-\frac{\pi}{N}} \mathrm{e}^{-\frac{2\pi}{N}(r+Nl)} \mathrm{e}^{i2\pi(r+Nl)\left(x+iy \right)} \mathrm{e}^{i2\pi(x+iy)}\\
&= \mathrm{e}^{-\frac{\pi}{2N}} \mathrm{e}^{i \pi x} (N/2)^{1/4} \mathrm{e}^{iN\pi \left(y+\frac{1}{N} \right) \left(x+iy+i\frac{1}{N} \right)} \sum_{l\in \mathbf{Z}} \mathrm{e}^{-\frac{\pi}{N}(r+Nl)^2} \mathrm{e}^{i2\pi(r+Nl)\left(x+iy+i\frac{1}{N} \right)}\\
&= \mathrm{e}^{-\frac{\pi}{2N}} \mathrm{e}^{i\pi x} \varphi^{(N)}_{0,r} (x, y+\frac{1}{N}).
\end{aligned}
\label{eq:A-3-4}
\end{equation}
Similarly, we obtain
\begin{equation}
\begin{aligned}
\bar{U}^{(N)}_{\tilde{r}r} \varphi^{(N)}_{0,\tilde{r}} (x,y)
&= \mathrm{e}^{-\frac{\pi}{2N}} \varphi^{(N)}_{0,r-1} (x,y)\\
&= \mathrm{e}^{-\frac{\pi}{2N}} \mathrm{e}^{-i\pi x} \varphi^{(N)}_{0,r} (x, y-\frac{1}{N}).
\end{aligned}
\label{eq:A-3-5}
\end{equation}
Using the above results, we find that
\begin{equation}
\begin{aligned}
V \circ T_{NN'}(\varphi^{(Q)}_{n,s})_{rr'} =& \mathrm{e}^{-\frac{\pi}{2N}} \left( \varphi^{(N')}_{0,r'} (x,y), \mathrm{e}^{i\pi y} \varphi^{(Q)}_{n,s} (x,y) \varphi^{(N)}_{0,r} (x-\frac{1}{N},y) \right)\\
& - \mathrm{e}^{-\frac{\pi}{2N'}} \left( \mathrm{e}^{-i\pi y} \varphi^{(N')}_{0,r'} (x+\frac{1}{N'},y), \varphi^{(Q)}_{n,s} (x,y) \varphi^{(N)}_{0,r} (x,y) \right),
\end{aligned}
\label{eq:A-3-6}
\end{equation}
and
\begin{equation}
\begin{aligned}
V^{\dagger} \circ ( V \circ T_{NN'}(\varphi^{(Q)}_{n,s}) )_{rr'} =& \mathrm{e}^{-\frac{\pi}{N}} \left( \varphi^{(N')}_{0,r'} (x,y), \varphi^{(Q)}_{n,s} (x,y) \varphi^{(N)}_{0,r} (x,y) \right)\\
& - \mathrm{e}^{-\frac{\pi}{2}\left(\frac{1}{N}+ \frac{1}{N'} \right)} \left( \varphi^{(N')}_{0,r'} (x-\frac{1}{N'},y), \varphi^{(Q)}_{n,s} (x,y) \varphi^{(N)}_{0,r} (x-\frac{1}{N},y) \right) \\
& - \mathrm{e}^{-\frac{\pi}{2}\left(\frac{1}{N}+ \frac{1}{N'} \right)} \left( \varphi^{(N')}_{0,r'} (x+\frac{1}{N'},y), \varphi^{(Q)}_{n,s} (x,y) \varphi^{(N)}_{0,r} (x+\frac{1}{N},y) \right) \\
& + \mathrm{e}^{-\frac{\pi}{N'}} \left( \varphi^{(N')}_{0,r'} (x,y), \varphi^{(Q)}_{n,s} (x,y) \varphi^{(N)}_{0,r} (x,y) \right) \\
=& \left( \mathrm{e}^{-\frac{\pi}{N}} + \mathrm{e}^{-\frac{\pi}{N'}} \right) \left( \varphi^{(N')}_{0,r'} (x,y), \varphi^{(Q)}_{n,s} (x,y) \varphi^{(N)}_{0,r} (x,y) \right)\\
& - \mathrm{e}^{-\frac{\pi}{2}\left(\frac{1}{N}+ \frac{1}{N'} \right)} \sum_{j=\pm 1} \left( \varphi^{(N')}_{0,r'} (x,y), \varphi^{(Q)}_{n,s} (x+\frac{j}{N'},y) \varphi^{(N)}_{0,r} (x-\frac{jQ}{NN'},y) \right),
\end{aligned}
\label{eq:A-3-7}
\end{equation}
where we used $(\mathrm{e}^{-i\pi y} \phi, \psi) = (\phi, \mathrm{e}^{i\pi y} \psi)$ and we also made a shift of the integral variable $x$ in the second equality.
We further rewrite
\begin{equation}
\begin{aligned}
\varphi^{(Q)}_{n,s} (x+\frac{j}{N'},y) \varphi^{(N)}_{0,r} (x-\frac{jQ}{NN'},y) &= ( \mathrm{e}^{\frac{j}{N'} \partial_x} \varphi^{(Q)}_{n,s} (x,y) )  ( \mathrm{e}^{-\frac{jQ}{NN'} \partial_x} \varphi^{(N)}_{0,r} (x,y) )\\
&= (\mathrm{e}^{\frac{j}{N'} (\partial_x + iQ\pi y)} \varphi^{(Q)}_{n,s} (x,y) ) ( \mathrm{e}^{-\frac{jQ}{NN'} (\partial_x +iN\pi y)} \varphi^{(N)}_{0,r} (x,y) )\\
&= ( \mathrm{e}^{\frac{j}{N'} D_x} \varphi^{(Q)}_{n,s} (x,y) )( \mathrm{e}^{-\frac{jQ}{NN'} D_x} \varphi^{(N)}_{0,r} (x,y) )
\end{aligned}\label{eq:A-3-8}
\end{equation}
in terms of the covariant derivatives $D_i$, which are 
given for fields with charge $N$ by
\begin{equation}
\begin{aligned}
&D_x = D_z + D_{\bar{z}} = \partial_x + i N \pi y \\
&D_y = i(D_z - D_{\bar{z}}) = \partial_y - i N \pi x .
\end{aligned}\label{eq:A-3-9}
\end{equation}
Therefore, we obtain
\begin{equation}
\begin{aligned}
V^{\dagger} \circ ( V \circ T_{NN'}(\varphi^{(Q)}_{n,s}))_{rr'} = & \left( \mathrm{e}^{-\frac{\pi}{N}} + \mathrm{e}^{-\frac{\pi}{N'}} \right) \left( \varphi^{(N')}_{0,r'}, \varphi^{(Q)}_{n,s} \varphi^{(N)}_{0,r}  \right)\\
& - \mathrm{e}^{-\frac{\pi}{2}\left(\frac{1}{N}+ \frac{1}{N'} \right)} \sum_{j=\pm 1} \left( \varphi^{(N')}_{0,r}, ( \mathrm{e}^{\frac{j}{N'} D_x} \varphi^{(Q)}_{n,s} )(\mathrm{e}^{-\frac{jQ}{NN'} D_x} \varphi^{(N)}_{0,r} ) \right).
\end{aligned}
\label{eq:A-3-10}
\end{equation}
We can repete the similar computation for $U$ and obtain
\begin{equation}
\begin{aligned}
U^{\dagger} \circ (U \circ T_{NN'}(\varphi^{(Q)}_{n,s}))_{rr'} = & \left( \mathrm{e}^{-\frac{\pi}{N}} + \mathrm{e}^{-\frac{\pi}{N'}} \right) \left( \varphi^{(N')}_{0,r'}, \varphi^{(Q)}_{n,s} \varphi^{(N)}_{0,r} \right)\\
& - \mathrm{e}^{-\frac{\pi}{2}\left(\frac{1}{N}+ \frac{1}{N'} \right)} \sum_{j=\pm 1} \left( \varphi^{(N')}_{0,r'}, ( \mathrm{e}^{\frac{j}{N'} D_y} \varphi^{(Q)}_{n,s} )( \mathrm{e}^{-\frac{jQ}{NN'} D_y} \varphi^{(N)}_{0,r} ) \right).
\end{aligned}
\label{eq:A-3-11}
\end{equation}
By summing \eqref{eq:A-3-10} and \eqref{eq:A-3-11},
we finally obtain \eqref{eq:4-2-84}.
%\begin{equation}
%\begin{aligned}
%\langle r | \hat{\Delta} (F) | r' \rangle =& N^2 \langle r | \left[ U^{\dagger} %\circ (U \circ F) + V^{\dagger} \circ (V \circ F) \right] | r' \rangle \\
%=& 2N^2 \left( \mathrm{e}^{-\pi/N} + \mathrm{e}^{-\pi/N'} \right) \langle r| T_{NN'}%(f^{(Q)s}_n) |r' \rangle \\
%& - N^2 \mathrm{e}^{-\frac{\pi}{2} \left( \frac{1}{N} + \frac{1}{N'} \right)}
%\sum_{i=x,y \, ; \, j=\pm1} \left( f^{(N')r'}_0 , ( \mathrm{e}^{\frac{j}{N'}D_i} f^{(Q)s}_n) \, (\mathrm{e}^{-\frac{jQ}{NN'}D_i} f^{(N)r}_0 ) \right).
%\end{aligned} .\label{eq:A-3-12}
%\end{equation}

%%%%%%%%%%%%%%%%%%%%%%%%%%%%%%%%%%%%%%%%%%%%%%%%%%%%%%%%%%%%%%%%%
\section{Laplacian for rectangular matrices and Hofstadter problem}
\label{Laplacian for rectangular matrices and Hofstadter problem}
In this appendix, we consider the exact eigenvalue problem 
of the matrix Laplacian (\ref{eq:4-2-82}) for rectangular matrices.
We show that the problem is equivalent to a special case of the Hofstadter problem \cite{1},
which we will review below.

The eigenvalue equation for $N \times N'$ matrices
is written as 
\begin{align}
\hat{\Delta} (F) = N^2 \left[ U^{\dagger} \circ (U \circ F) + V^{\dagger} \circ (V \circ F) \right] =EF.
\label{eq:4-2-87}
\end{align}
In terms of the matrix elements of $F$, this is equivalent to
\begin{align}
F_{r+1,r'+1} + 2 \cos \left( 2 \pi \left( \frac{r}{N} -\frac{r'}{N'} \right)\right)F_{r,r'} + F_{r-1,r'-1} = \tilde{E} F_{r,r'},
\label{eq:4-2-88}
\end{align}
where 
$\tilde{E}$ is given by
\begin{align}
\tilde{E}=4 \cosh \left( \frac{Q\pi}{2NN'} \right) - \frac{E}{N^2} \mathrm{e}^{\frac{\pi}{2}\left( \frac{1}{N} + \frac{1}{N'} \right)}.
\label{eq:4-2-89}
\end{align}
The periodic structure of (\ref{eq:4-2-88}) enables us to 
extend the range of indices as $F_{r+N, r'+N'} = F_{r, r'}$.
With this notation, assuming that $N$ and $N'$ are coprime,
we relabel the matrix elements as
\begin{equation}
F_r := F_{r,r}
\label{eq:4-2-90}
\end{equation}
for $r = 0,1, \cdots , NN'-1$.
In this notation, (\ref{eq:4-2-88}) reduces to
\begin{equation}
F_{r+1} + 2 \cos \left( \frac{2 Q \pi r}{NN'}\right) F_{r} + F_{r-1} = \tilde{E} F_{r} 
\label{eq:4-2-91}
\end{equation}
for $r = 0,1, \cdots , NN'-1$,  
where $F_{-1} := F_{NN'-1}$ and $F_{NN'} := F_{0}$.
This is also equivalent to the following eigenvalue problem:
\begin{equation}
H \vec{F} = \tilde{E} \vec{F},
\label{eq:4-2-92}
\end{equation}
where
\begin{align}
H &= (C^{(NN')})^{Q} + (C^{(NN') \dagger})^{Q} + S^{(NN')} + S^{(NN') \dagger} \label{eq:4-2-93}\\
\vec{F} &= (F_0, F_1, \cdots, F_{NN'-1})^{\mathrm{T}}. 
\label{eq:4-2-94}
\end{align}
The eigenvalue problem of $H$ is what is known as the Hofstadter problem \cite{1}.
Finding an exact solution to this problem is still an open problem, though 
some numerical analyses have been done \cite{1} and revealed a 
fractal structure of the spectrum, known as a Hofstadter butterfly.

It is interesting that
the same Hofstadter problem also arises in a system of 
tight-binding Bloch electrons under a constant magnetic flux in 
a periodic two-dimenisonal surface, which has the following Hamiltonian:
\begin{equation}
H = -t \sum_{i=0}^{q-1} \sum_{\vec{k}} \sum_{\sigma= \uparrow \downarrow}  \omega_{i} (\vec{k}) d^{\dagger}_{i,\sigma} ( \vec{k} ) d_{i, \sigma}   ( \vec{k} ),
\label{eq:4-2-95}
\end{equation}
where $t$ is the hopping parameter, $q$ is the number of lattice sites,
$d_{i,\sigma}(\vec{k})$ is a creation and annihilation operator for the wave number $\vec{k}$ and spin $\sigma$, satisfying anti-commutation relations
$\{d_{i,\sigma}(\vec{k}) \, , \, d^{\dagger}_{j,\sigma'}(\vec{k} ') \}=\delta_{ij} \delta_{\sigma \sigma'} \delta_{\vec{k} \vec{k} '}$.
The eigenvalue $\omega_{i} (\vec{k})$ is obtained by solving the 
eigenvalue problem of
\begin{equation}
H(\vec{k})=  \left(
\begin{array} {cccccc}
      2  \cos k_{2} \,\,\,&                         1           &       0        &   \cdots    &   \,\, 0 \,\,        &    e^{-iqk_{1} }    \\
           1                   &  2\cos(k_{2}-2\pi \phi)    &             1   &                &  \,\,\,\,             &          0             \\
           0                  &                       1             &     \ddots  &    \ddots    &     \,\,\,\,          &       \vdots         \\
           \vdots           &                                      & \ddots        &    \ddots  & \,\,  \ddots \,\, &           0             \\
           0                  &                                      &                 &    \ddots    & \,\, \ddots \,\,   &           1             \\
      e^{iqk_{1}}          &                     0               &   \cdots    &       0         &   \,\,1 \,\,          &  2\cos(k_{2} -  2\pi \phi (q-1) )
\end{array}
\right),
\label{eq:4-2-96}
\end{equation}
where $\phi=\frac{p}{q}$ is the
the strength of $U(1)$ flux per unit plaquette and $p$ is the Chern number.
Here, we assumed that $p$ and $q$ are coprime for simplicity.
Readers may refer to \cite{Kohmoto} for the derivations of the Hamiltonian \eqref{eq:4-2-95} and the matrix \eqref{eq:4-2-96}.
If we put $\phi = \frac{Q}{NN'}, \,\, q = NN'$ and $\vec{k} = 0$, the matrix 
\eqref{eq:4-2-96} reduces to the Hamiltonian \eqref{eq:4-2-93} for the 
matrix Laplacian.

The spectrum of \eqref{eq:4-2-93} or \eqref{eq:4-2-96}
has been studied numerically. In the large-$q$ limit, 
it is shown that the spectrum of \eqref{eq:4-2-93}
indeed coincides with the Landau level \cite{2}, 
and this is consistent with our result.

%%%%%%%%%%%%%%%%%%%%%%%%%%%%%%%%%%%%%%%%%%%%%%%%
\section{Evaluating the norm of $\psi^-$}
\label{norm on higher genus}

In this appendix, we show that the norm of 
$\psi^-$, which is given by 
(\ref{Dirac zero mode on higher genus 1})
and (\ref{automorphic form}), does not converge.

First, for the inner product (\ref{inner product of spinors}),
the norm of $\psi^-$ can be rewritten as
\als
{
	\label{norm on higher genus 1}
	\|\psi^-\|^2
	&=
	\sum_{\gamma,\gamma'}\int_\mathcal{M} \omega\,(1-|z|^2)^{1-N}
	\left(\frac{d\gamma(z)}{dz}
	\frac{d\bar\gamma'(z)}{d\bar z}\right)^{(1-N)/2}
	\bar f^-(\bar\gamma(z))f^-(\bar\gamma'(z))\\
	&=
	\sum_{\gamma,\eta}\int_\mathcal{M} \omega\,(1-|z|^2)^{1-N}
	\left(\frac{d\gamma(z)}{dz}\frac{d(\bar\eta\bar\gamma)(z)}{d\bar z}\right)^{(1-N)/2}
	\bar f^-(\bar\gamma(z))f^-((\bar\eta\bar\gamma)(z))\\
	&=
	\sum_{\gamma,\eta}\int_{\gamma^{-1}(\mathcal{M})}
	\omega\,(1-|\gamma^{-1}(w)|^2)^{1-N}
	\left(\frac{dw}{d\gamma^{-1}(w)}
	\frac{d\bar\eta(\bar w)}{d\bar\gamma^{-1}(w)}\right)^{(1-N)/2}
	\bar f^-(\bar w)f^-(\bar\eta(\bar w))\\
	&=
	\sum_{\gamma,\eta}\int_{\gamma^{-1}(\mathcal{M})}
	\omega\,(1-|w|^2)^{1-N}
	\left(\frac{d\bar\eta(\bar w)}{d\bar w}\right)^{(1-N)/2}
	\bar f^-(\bar w)f^-(\bar\eta(\bar w)).
}
To obtain the second equality we changed the dummy 
variable from $\gamma'$ to $\eta=\gamma'\gamma^{-1}$ and
to obtain the third equality, we changed the integral variable 
by $w=\gamma(z)$. Note that $\omega$ is invariant 
under actions of $\Gamma$. To obtain the last equality, 
we used the fact that for any $\gamma\in\Gamma$, the relation
\beq
	\label{relation for SU(1,1) action}
	1-|\gamma(z)|^2
	=
	\left|\frac{d\gamma(z)}{dz}\right|(1-|z|^2)
\eeq
holds, so that
\als
{
	(1-|\gamma^{-1}(w)|^2)^{1-N}
	\left(\frac{dw}{d\gamma^{-1}(w)}
	\frac{d\bar\eta(\bar w)}{d\bar\gamma^{-1}(w)}\right)^{(1-N)/2}
	=
	(1-|w|^2)^{1-N}
	\left(\frac{d\bar\eta(\bar w)}{d\bar w}\right)^{(1-N)/2}.
}
In the last line of (\ref{norm on higher genus 1}), 
we can use the relation
$\sum_\gamma\int_{\gamma^{-1}(\mathcal{M})}=\int_{D^2}$.
Hence, from
$\omega=idw\wedge d\bar w/(1-|w|^2)^2$, we obtain
\als
{
	\|\psi^-\|^2
	&=
	\sum_{\eta}\int^1_0d|w|^2(1-|w|^2)^{-1-N}
	\int^{2\pi}_0d(\text{arg}w)
	\left(\frac{d\bar\eta(\bar w)}{d\bar w}\right)^{(1-N)/2}
	\bar f^-(\bar w)f^-(\bar\eta(\bar w)).
}
This shows that for $N\geq1$, the integration of $|w|^2$
does not converge.
Thus, $	\|\psi^-\|^2$ is not convergent for $N\geq 1$.

%%%%%%%%%%%%%%%%%%%%%%%%%%%%%%%%%%%%%%%%%%%%%%%%
\section{Bergman kernel on disk}
\label{Bergman kernel on disk}

In this appendix, we construct a Bergman kernel on the 
Poincar$\acute{\text{e}}$ disk $D^2$.
See \cite{Klimek:1992a} for more details.

On the Poincar$\acute{\text{e}}$ disk, 
an orthonormal basis of the Dirac zero mode is given by
\beq
	\psi_n(z,\bar z)
	=
	(1-|z|^2)^{(N+1)/2}\left(\frac{N}{2\pi}\right)^{1/2}
	\binom{N+n}{N}^{1/2}z^n.
\eeq
Here, $n=1,2,\ldots,\infty$, so the dimension of $\text{Ker}D$ is 
infinity. This comes from the noncompactness of 
$D^2$. We can check the orthonormality as follows.
\als
{
	(\psi_n,\psi_m)
	&=
	\frac{N}{2\pi}
	\binom{N+n}{N}^{1/2}\binom{N+m}{N}^{1/2}
	\int^1_0d|z|^2(1-|z|^2)^{N-1}\int^{2\pi}_0d(\text{arg}z)\bar z^nz^m\\
	&=
	\delta_{nm}N\binom{N+n}{n}
	\int^1_0d|z|^2(1-|z|^2)^{N-1}|z|^{2n}\\
	&=\delta_{nm}N\binom{N+n}{n}
	\frac{\Gamma(N)\Gamma(n+1)}{\Gamma(N+n+1)}\\
	&=
	\delta_{nm}.
}
By using the generalized binomial theorem
$(1-x)^{-(N+1)}=\sum^\infty_{n=0}\binom{N+n}{N}x^n$,
we find that the Bergman kernel is given by
\als
{
	K^{(N)}(z,w)
	&=
	\frac{N}{2\pi}(1-|z|^2)^{(N+1)/2}(1-|w|^2)^{(N+1)/2}
	(1-\bar zw)^{-(N+1)}.
}

\end{appendix}

\end{document}